\documentclass{aa}  
\usepackage[varg]{txfonts}

\bibpunct{(}{)}{;}{a}{}{,} 
\usepackage{graphicx}

\begin{document} 

\title{Kelvin-Helmholtz instability and heating in oscillating loops perturbed by power-law transverse wave drivers}

\author{%
{Konstantinos Karampelas}\inst{\ref{aff:CmPA}} \orcid{0000-0001-5507-1891}
\and {Tom Van Doorsselaere}\inst{\ref{aff:CmPA}} \orcid{0000-0001-9628-4113}
\and {Mingzhe Guo}\inst{\ref{aff:CmPA}} \orcid{0000-0003-4956-6040}
\and {Timothy Duckenfield}\inst{\ref{aff:northumbria}} \orcid{0000-0003-3306-4978}
\and {Gabriel Pelouze}\inst{\ref{aff:LIFA}}\orcid{0000-0002-0397-2214}
}

\institute{%
\label{aff:CmPA}{Centre for mathematical Plasma Astrophysics, Department of Mathematics, KU Leuven, Celestijnenlaan 200B, 3001 Leuven, Belgium.}\\ \email{kostas.karampelas@kuleuven.be}
\and
\label{aff:northumbria}{Department of Mathematics, Physics and Electrical Engineering, Northumbria University, Newcastle upon Tyne, NE1 8ST, UK}
\and
\label{aff:LIFA}{LifeWatch ERIC, Virtual Lab \& Innovation Center (VLIC), University of Amsterdam, Science Park 904, 1098 XH, Amsterdam, The Netherlands
}
}

\date{Received 2023-12-15; Accepted 2024-05-08}
 
\abstract
{Instabilities in oscillating loops are believed to be essential for dissipating the wave energy and heating the solar coronal plasma.}
{Our aim is to study the development of the Kelvin-Helmholtz (KH) instability in an oscillating loop that is driven by random footpoint motions.}
{Using the PLUTO code, we performed 3D simulations of a straight gravitationally stratified flux tube. The loop footpoints are embedded in chromospheric plasma, in the presence of thermal conduction and an artificially broadened transition region. Using drivers with a power-law spectrum, one with  a red noise spectrum and one with the low-frequency part subtracted, we excited standing oscillations and the KH instability in our loops, after one-and-a-half periods of the oscillation.}
{We see that our broadband drivers lead to fully deformed, turbulent loop cross-sections over the entire coronal part of the loop due to the spatially extended KH instability. The low RMS velocity of our driver without the low-frequency components supports the working hypothesis that the KH instability can easily manifest in oscillating coronal loops. We report for the first time in driven transverse oscillations of loops the apparent propagation of density perturbations due to the onset of the  KH instability, from the apex towards the footpoints. Both drivers input sufficient energy to drive enthalpy and mass flux fluctuations along the loop, while also causing heating near the driven footpoint of the oscillating loop, which becomes more prominent when a low-frequency component is included in the velocity driver. Finally, our power-law driver with the low-frequency component provides a RMS input Poynting flux of the same order as the radiative losses of the quiet-Sun corona, giving us promising prospects for the contribution of decayless oscillations in coronal heating.}
{}

\keywords{magnetohydrodynamics - solar coronal seismology - solar coronal waves - magnetohydrodynamical simulations}

\titlerunning{KH instability in loops perturbed by power-law wave drivers}
\authorrunning{Karampelas et al.}

\maketitle
\nolinenumbers
\section{Introduction} \label{sec:introduction}

When studying the physics of the solar atmosphere, it quickly becomes clear that plasma is highly organised in pronounced magnetic structures, such as coronal loops, observed in the extreme ultraviolet (EUV) and soft X-rays \citep[see][for a review]{Reale2014}. This structuring is crucial for both the manifestation and propagation of magnetohydrodynamic (MHD) waves \citep[e.g.][]{edwin1983wave} and for energy dissipation. This is further supported by the abundance of observations of waves in the solar atmosphere. Of particular interest are the transverse waves \citep[see][for a review]{NakariakovEtAl2021}. Since their first detection \citep{aschwanden1999,nakariakov1999}, the ubiquitous nature of these waves has been proven through multiple observations from different instruments \citep[e.g.][]{tomczyk2007,mcintosh2011,tian2012,wang2012, Morton2015NatCo...6.7813M}. Focusing on standing kink oscillations \citep{tvd2008detection} in coronal loops, there are two different regimes to be considered. The first is that of the large-amplitude fast-decaying oscillations excited by external energetic phenomena \citep[e.g.][]{nakariakov1999,Nechaeva2019ApJS}. The second consists of oscillations of lower amplitude that do not visibly decay for many periods and are descriptively called decayless oscillations \citep[e.g.][]{nistico2013,anfinogentov2013}. They have been shown to be ubiquitous in the solar corona \citep[e.g.][]{anfinogentov2015,ZhongSihui2022MNRAS.516.5989Z}, and have also been observed in the lower solar atmosphere \citep{Petrova2023ApJ...946...36P,ShrivastavArpitKumar2024A&A...685A..36S,GaoYuhang2024A&A...681L...4G} and in coronal bright points \citep{GaoYuhang2022ApJ...930...55G}. 

The transport of energy from the solar photosphere to the atmosphere and its dissipation to support its multi-million degree temperature corona are topics that are  still not fully understood to this day. The proposed heating mechanisms are usually separated into two categories, by comparing their timescales to the Alfv\'{e}n transit time, the direct current (DC) and alternating current (AC) mechanisms \citep[see reviews][]{demoortel2015review,tvd2020coronalSSRv..216..140V}. Transverse oscillations are expected to lose energy through resonant absorption \citep[e.g.][]{goossens2002coronal}, phase mixing \citep[e.g.][]{heyvaerts1983,pagano2017}, and the damping induced by the  Kelvin-Helmholtz (KH) instability   \citep[e.g.][]{terradas2008,magyar2016damping,tvd2021ApJ...910...58V}. Unless we consider decayless oscillations to be Line-Of-Sight effects due to the development of instabilities in decaying oscillations \citep{antolin2016}, only continuous energy injection can sustain their amplitudes. That makes them  important candidates for an AC type of coronal heating mechanism. The exact mechanism exciting decayless oscillations has not yet been identified, but they are most often  modelled numerically as standing waves driven by monoperiodic \citep[e.g.][]{karampelas2017,mingzhe2019} and broadband footpoint drivers \citep{afanasyev2020decayless,Ruderman2021MNRAS.501.3017R,Ruderman2021SoPh..296..124R,Howson2023Physi...5..140H,Karampelas2024A&A...681L...6K}, by external flows via vortex shedding \citep{nakariakov2009,karampelas2021ApJ...908L...7K}, and as  self-oscillatory processes \citep{nakariakov2016,karampelas2020ApJ}.

In the last few years simulations of driven decayless oscillations have shown that they carry enough energy to counterbalance the optically thin radiative losses in the corona \citep[e.g.][]{mijie2021ApJL,DeMoortel2022ApJ...941...85D}. However, phase mixing from transverse waves generated by footpoint broadband drivers is unlikely to support coronal temperatures via wave heating \citep[e.g.][]{Pagano2019,Pagano2020A&A...643A..73P}. Staying with broadband drivers, \citet{Howson2023Physi...5..140H} have shown that linearly  polarised drivers are the most efficient in terms of energy input into driven loop oscillations, as expected from the linear wave theory. In the same study, the development of the KH instability was reported for broadband drivers of various polarisations. However, only the resonant linearly polarised driver led to spatially extended KH instability eddies and fully deformed loops, as had been shown for the first time in \citet{karampelas2018fd}. The KH instability-induced turbulence in transversely \citep{karampelas2019} and torsionally oscillating loops \citep{DiazSuarez2021A&A...648A..22D,DiazSuarez2022A&A...665A.113D,DiazSuarez2023A&A...670C...4D...Corrigendum} transfers energy to smaller scales, leading to dissipation and potentially to plasma heating. The coupling with the chromosphere, and its response to heating based on resonant absorption of Alfv\'{e}n waves \citep{ofman1998,VanDamme2020A&A...635A.174V} and from non-linear  standing  kink waves \citep{mingzhe2023ApJ...949L...1G} have also been considered. At the same time, it has been argued that energy release in the corona is more likely to be driven by low-frequency motions \citep{Howson2022A&A...661A.144H}.  It is therefore important to include a more realistic driver with a strong low frequency (or `DC') component when studying the response of the lower atmosphere to the dynamics and heating of coronal loops.

In this study we  model driven transverse waves, generated by random motions, in a gravitationally stratified loop. In our recent study \citep[][]{Karampelas2024A&A...681L...6K}, we  show that this model can lead to the development of transverse oscillations resembling the recently observed decayless oscillations. Here we  focus our study on the development of the  KH instability and the energy evolution our models. In Section \ref{sec:setup} we describe our  set-up and the methodology used to construct it. The results of our simulations are described in detail in Section \ref{sec:results}. Finally, a thorough discussion of the important points of this study is presented in Section \ref{sec:discussions}.

\section{Numerical set-up} \label{sec:setup}
\begin{figure*}
    \centering
    \resizebox{\hsize}{!}{
    \includegraphics[trim={0.cm 0.cm 0.cm 0.cm},clip,scale=0.55]{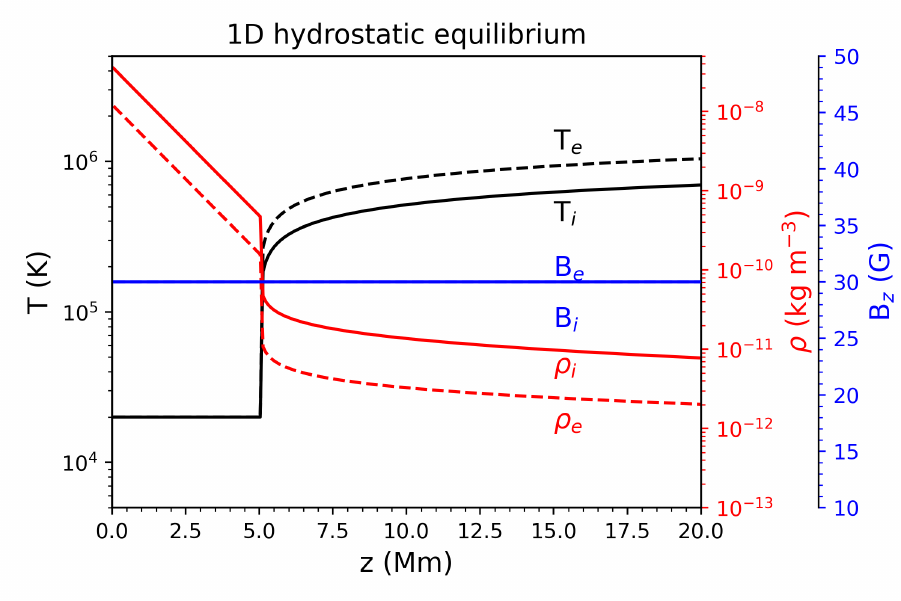}
    \includegraphics[trim={0.cm 0.cm 0.cm 0.cm},clip,scale=0.55]{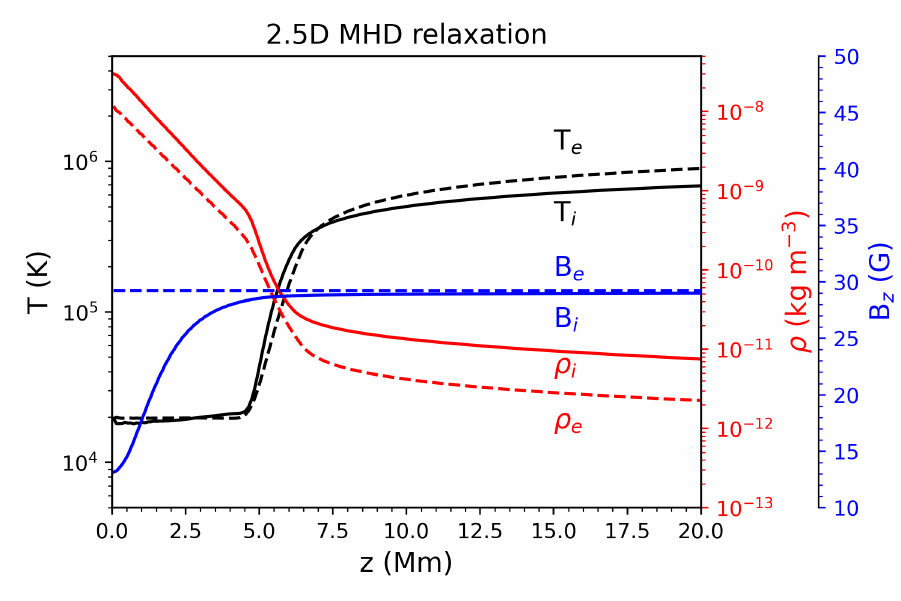}
    }
    \caption{Temperature, density and $B_z$ magnetic field profiles along the vertical direction, for $z\in \left[ 0,20 \right]$\,Mm. The left panel shows the results of solving the 1D equation of hydrostatic equilibrium for the inside (solid line) and outside (dashed line) of the flux tube. The right panel shows the same, but at the end of the 2.5D MHD relaxation. The subscripts refer to the values external ($e$) and internal ($i$) to the loop.}
    \label{fig:hydrostatic}
\end{figure*}

\begin{figure}
    \centering
    \resizebox{\hsize}{!}{
    \includegraphics[trim={0.2cm 0.2cm 0.cm 0.cm},clip,scale=0.5]{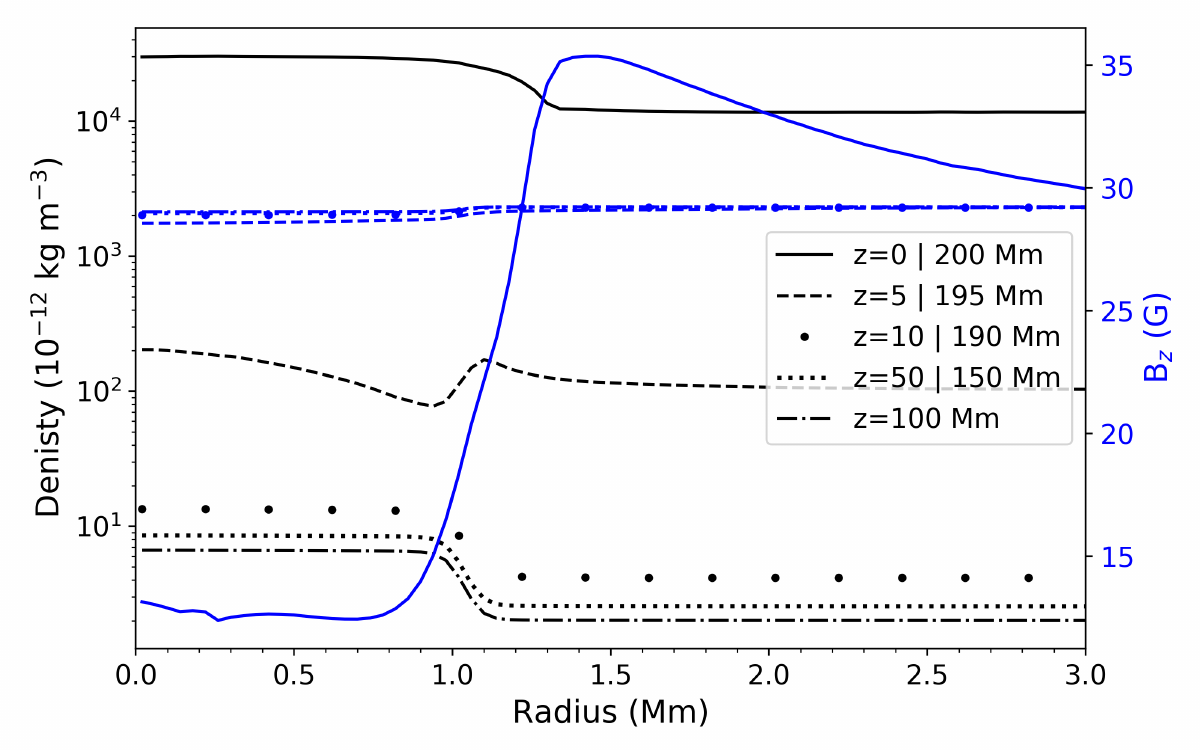}}
    \caption{Radial density and $B_z$ magnetic field profiles at different heights $z$ for the flux tube after the 2.5D MHD relaxation.}
    \label{fig:DTprofile}
\end{figure}

\begin{figure*}
    \centering
    \includegraphics[trim={0.cm 0.8cm 0.cm 0.cm},clip,scale=0.45]{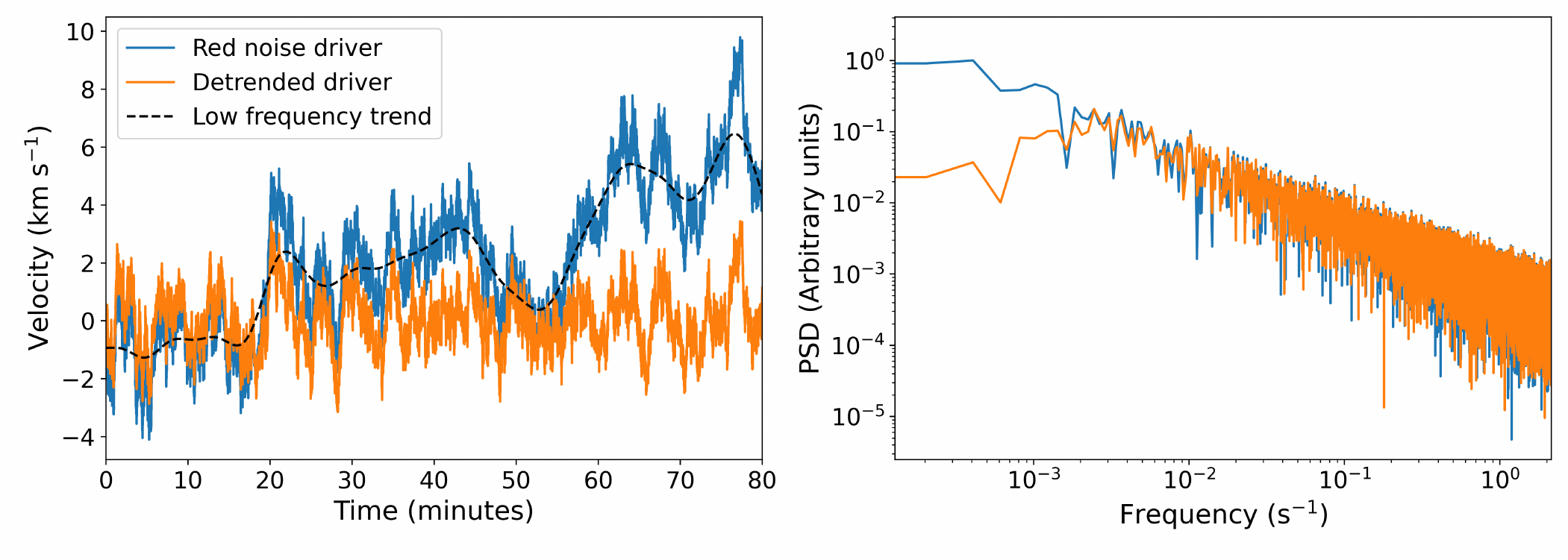}
    \caption{Velocity profiles (left panel) and spectra (right panel) for our drivers. The blue curve corresponds to the red noise signal with a power-law ($\propto f^{-1.66})$ spectrum, and the orange curve to the detrended signal with spectrum $S\propto f^{-1.66}$ for $f\gtrsim 2$\,mHz and reduced power at the lower frequencies. The black dashed line  shows the background trend for the original velocity signal.}
    \label{fig:Vprofile}
\end{figure*}

\begin{figure*}
    \centering
    \includegraphics[trim={0.cm 0.cm 0.cm 0.8cm},clip,scale=0.48]{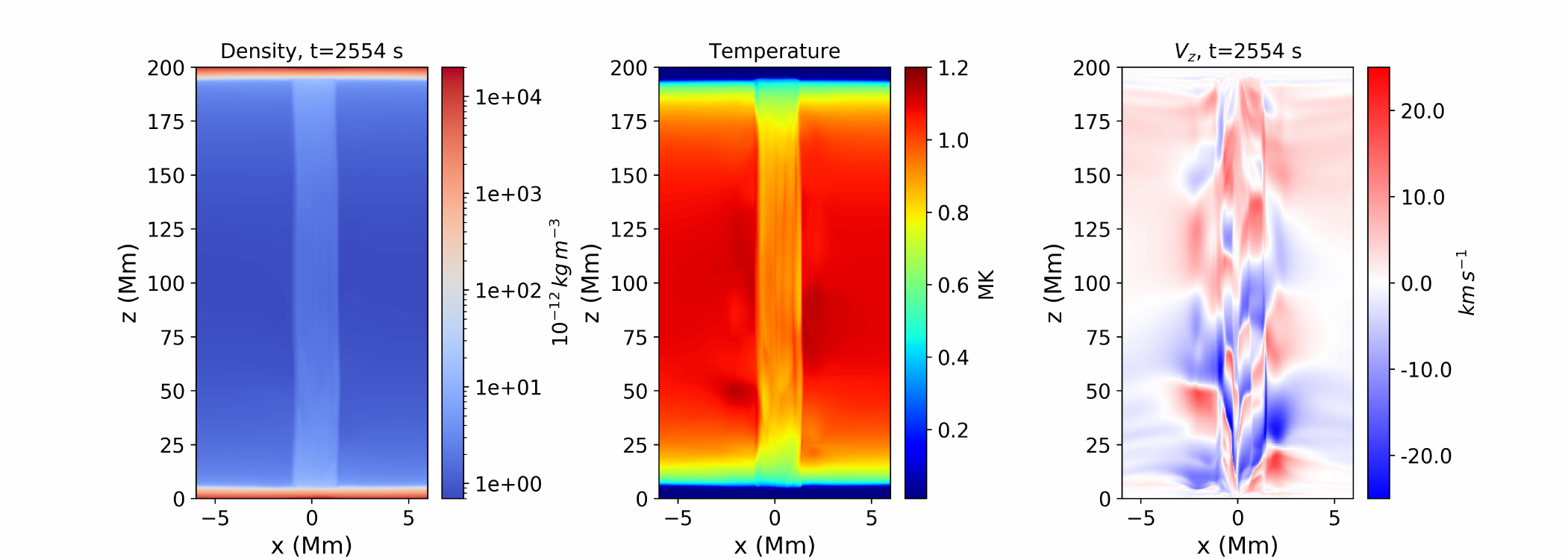}
    \includegraphics[trim={0.cm 0.cm 0.cm 0.8cm},clip,scale=0.48]{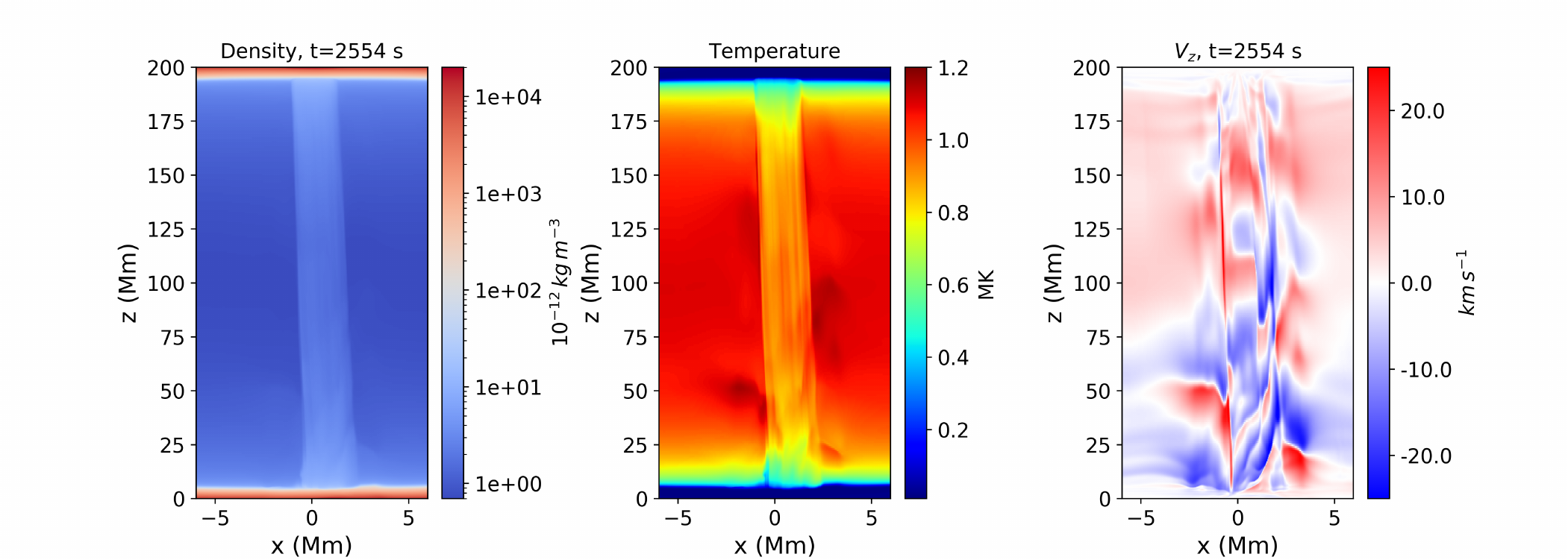}
    \caption{Integrated density (left) and temperature (middle) along the $y$-axis, and the $v_z$ velocity at the $y=0$ plane (right). The top and bottom panels show the domain at the last snapshots for the simulations with the detrended and red noise driver, respectively, at time $t=2554$\,s. Animations of the top and bottom panels are included in the online version of this manuscript.}
    \label{fig:Comp}
\end{figure*}

\begin{figure}
    \centering
    \includegraphics[trim={.5cm 0.cm 0.cm 0.cm},clip,scale=0.45]{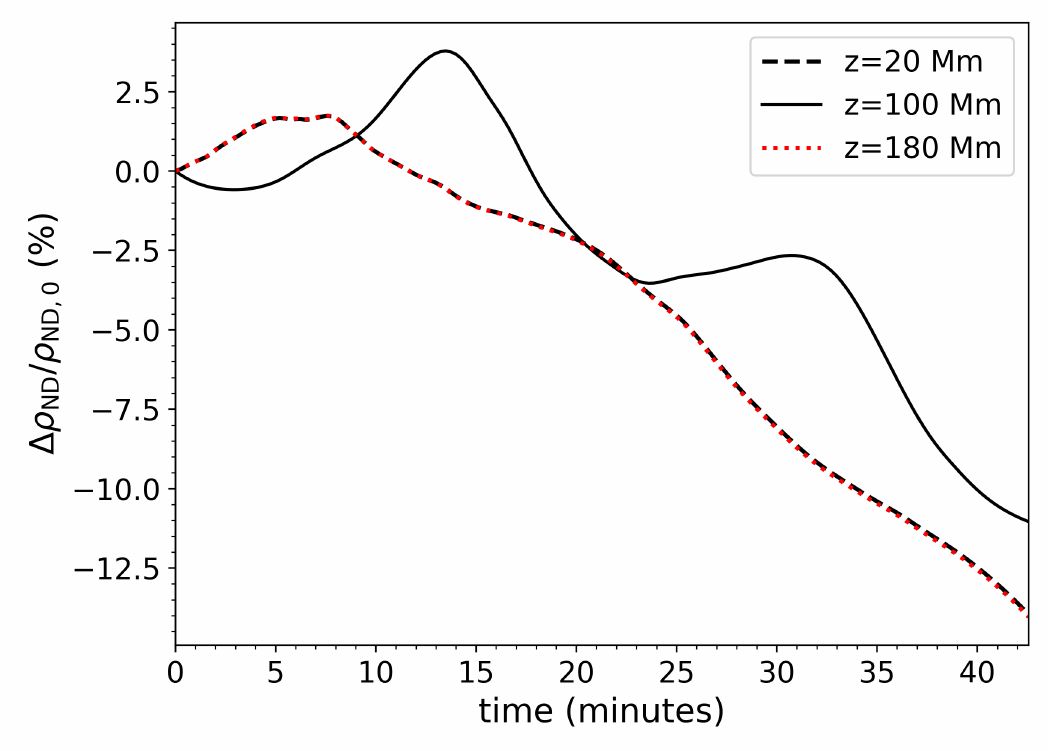}
    \caption{Profiles at different heights of the mean density over time, for the loop without a driver (see $\mathrm{ND}$ subscript for non-driven). The three different heights  are $z=100$\,Mm (apex), $z=20$\,Mm and $z=180$\, Mm. The latter two showcase the symmetry of the non-driven case with respect to the apex.}
    \label{fig:Density}
\end{figure}

\begin{figure}
    \centering
    \includegraphics[trim={.5cm 0.cm 0.cm 0.cm},clip,scale=0.45]{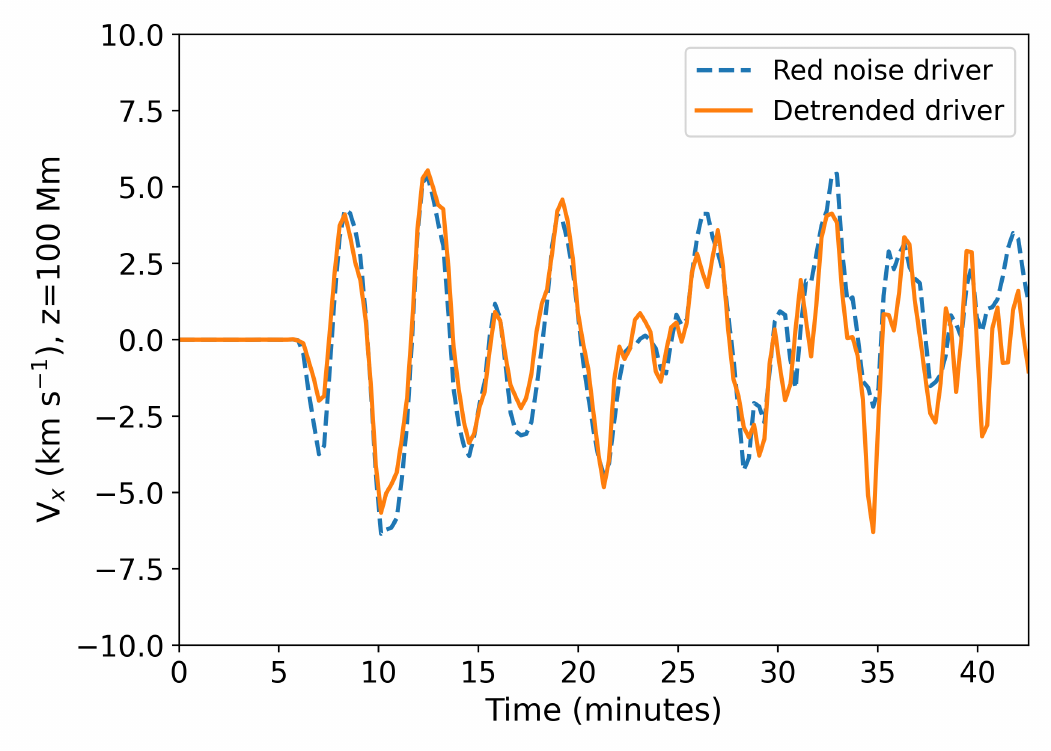}
    \caption{Velocity profiles at $z=100$\,Mm (apex), calculated from the centre of mass displacement along the direction of oscillation (x direction) for the two oscillating loops. The blue dashed line and the orange solid line correspond to the transverse oscillations excited by the red noise driver and detrended driver, respectively.}
    \label{fig:Vxapex}
\end{figure}

\begin{figure*}
    \centering
    \resizebox{\hsize}{!}{
    \includegraphics[trim={0.cm 0.cm 3.4cm 0.cm},clip,scale=0.48]{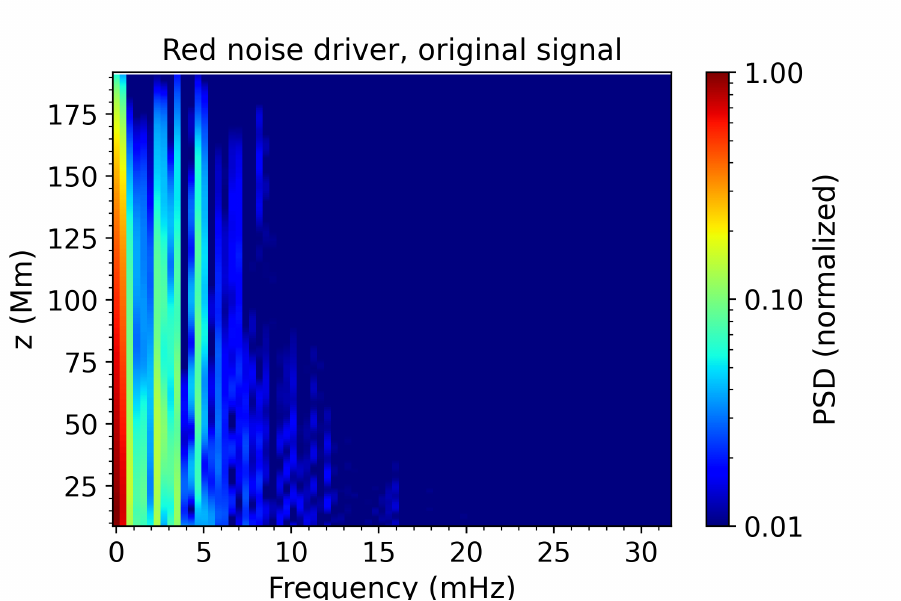}
    \includegraphics[trim={1.8cm 0.cm 3.4cm 0.cm},clip,scale=0.48]{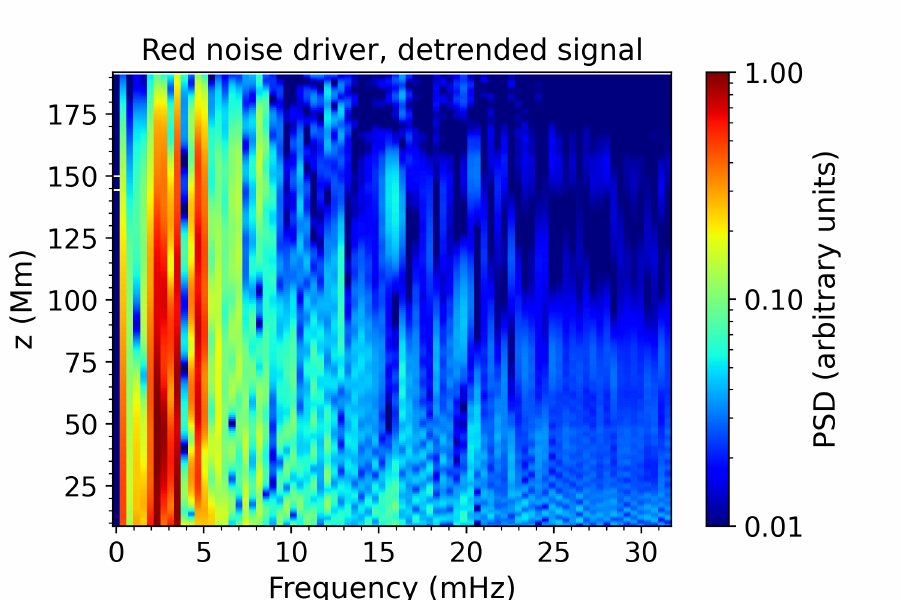}
    \includegraphics[trim={1.8cm 0.cm 3.4cm 0.cm},clip,scale=0.48]{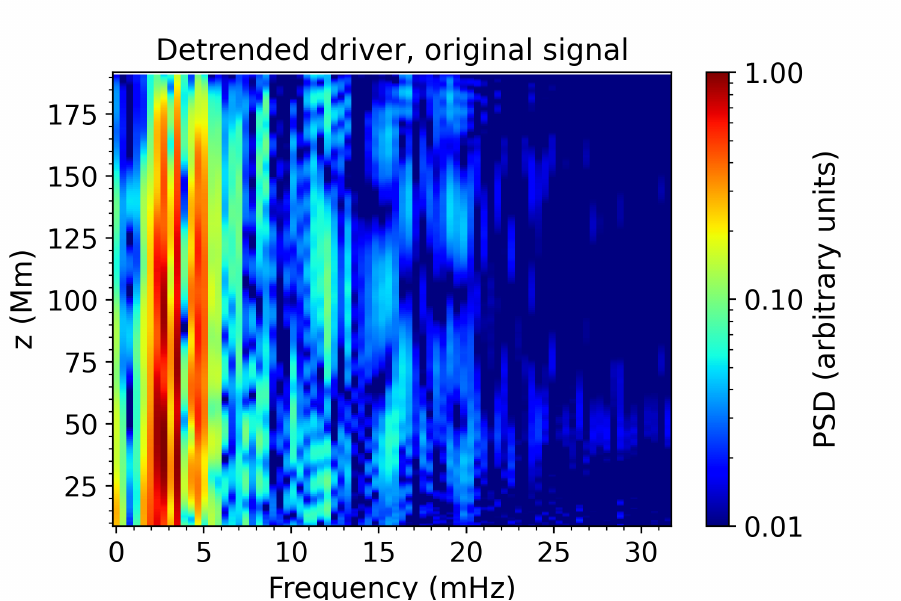}
    }
    \caption{Fourier power spectral density (PSD) profiles of the centre of mass displacement, along the coronal part of our loop. The left panel corresponds to the loop perturbed by the red noise driver. The middle panel corresponds to the detrended version of the displacement signal generated by the red noise driver. The right panel corresponds to the loop perturbed by the detrended driver. The PSD resolution along $z$ equals the numerical resolution of our set-ups.}
    \label{fig:Fourier}
\end{figure*}

\begin{figure*}
    \centering
    \includegraphics[trim={0.cm 0.cm 0.cm 0.8cm},clip,scale=0.48]{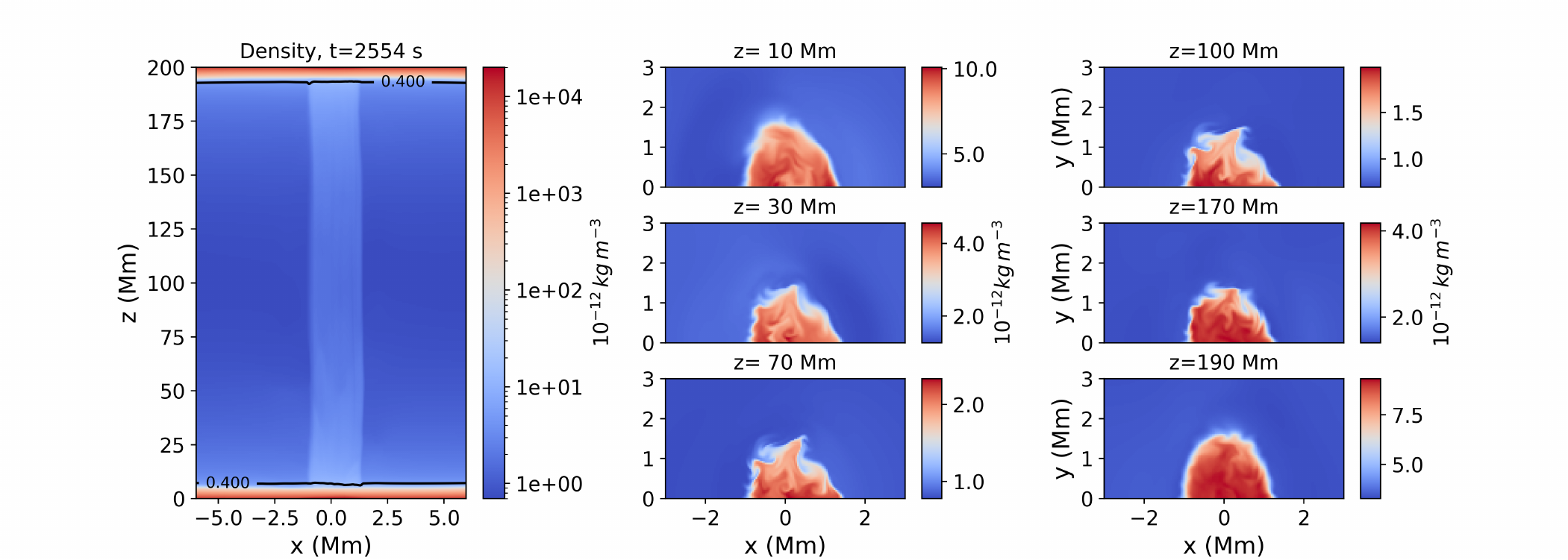}
    \includegraphics[trim={0.cm 0.cm 0.cm 0.8cm},clip,scale=0.48]{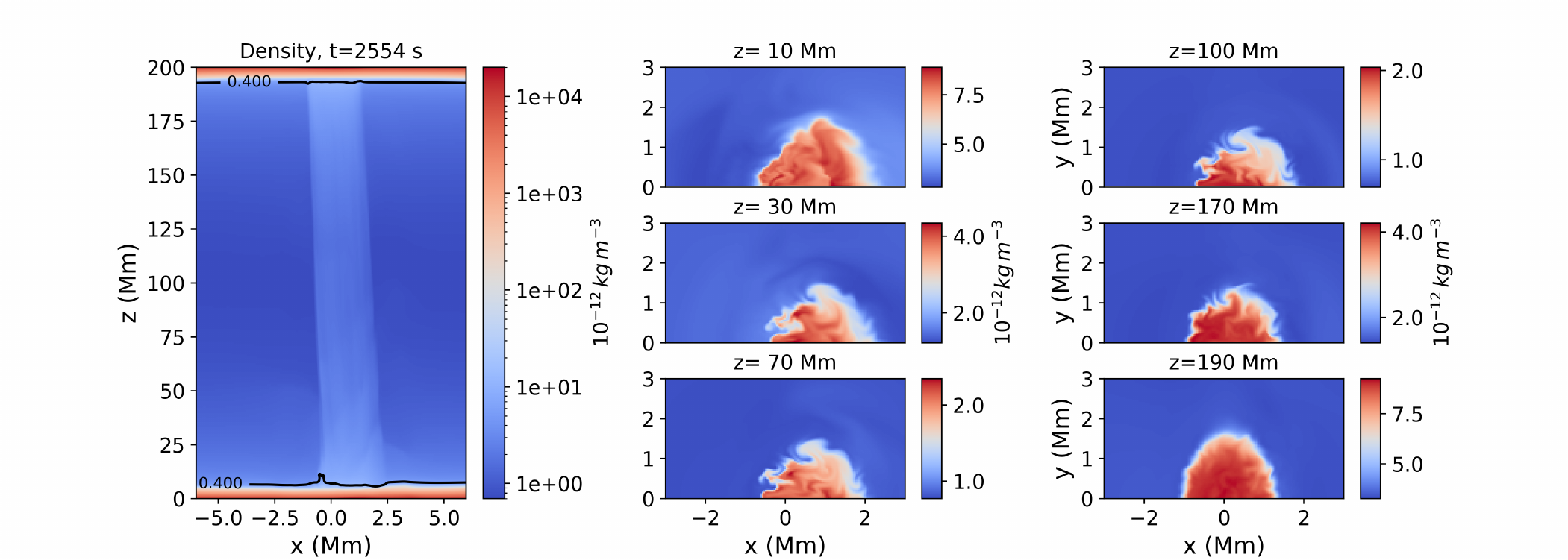}
    \caption{Integrated density along the $y$-axis and cross-sections of our loop at different heights, shown at time $t=2552$\,s. The top panels are for the detrended driver and the bottom panels are for the red noise driver. The black contour lines show the heights where the integrated temperature along the $y$-axis gets the value $T=0.4$\,MK. Animations of the two panels are included in the online version of this manuscript.}
    \label{fig:KHI}
\end{figure*}

\begin{figure*}
    \centering
    \resizebox{\hsize}{!}{
    \includegraphics[trim={0.cm 0.cm 4.8cm 0.cm},clip,scale=0.38]{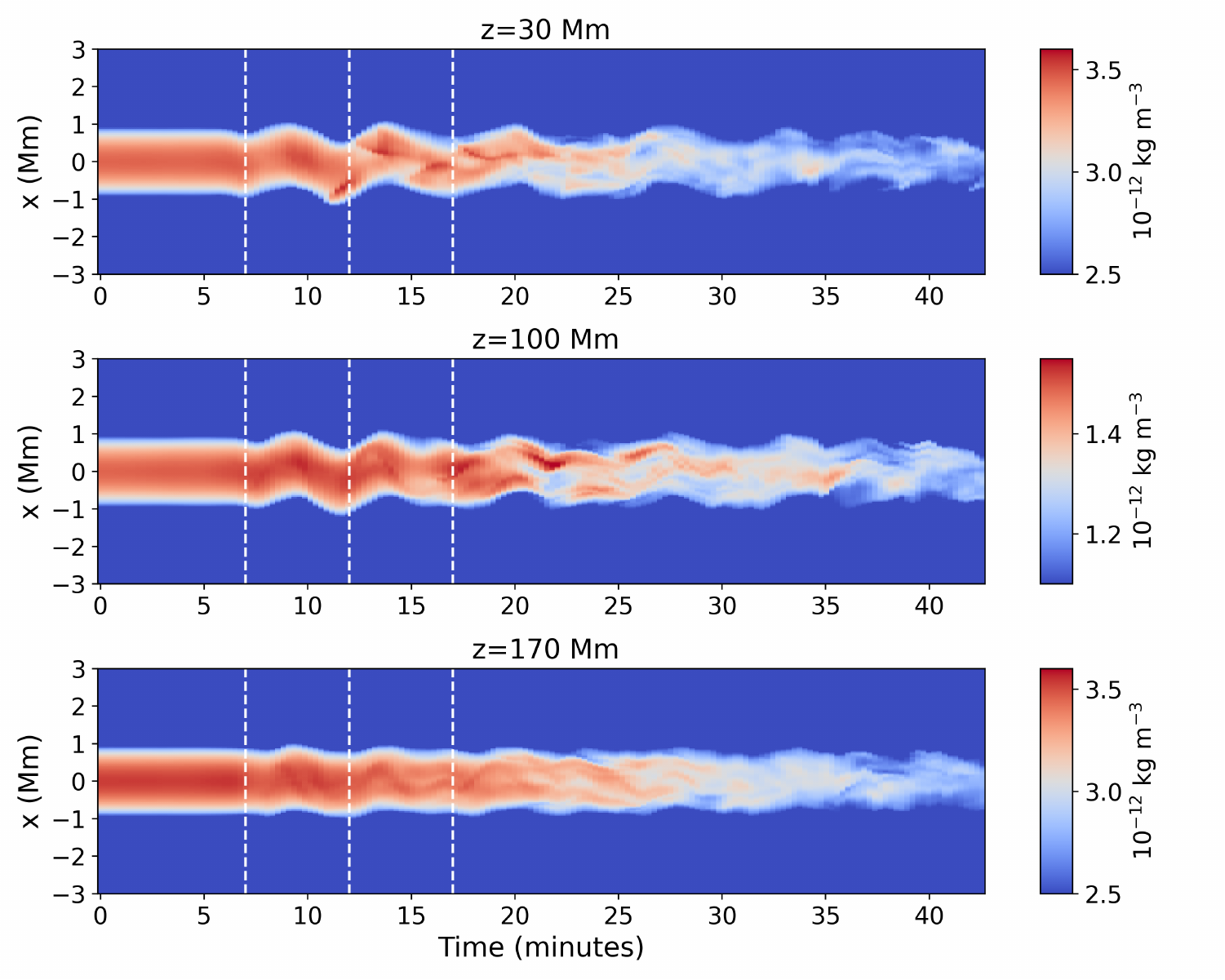}
    \includegraphics[trim={1.95cm 0.cm 0.cm 0.cm},clip,scale=0.38]{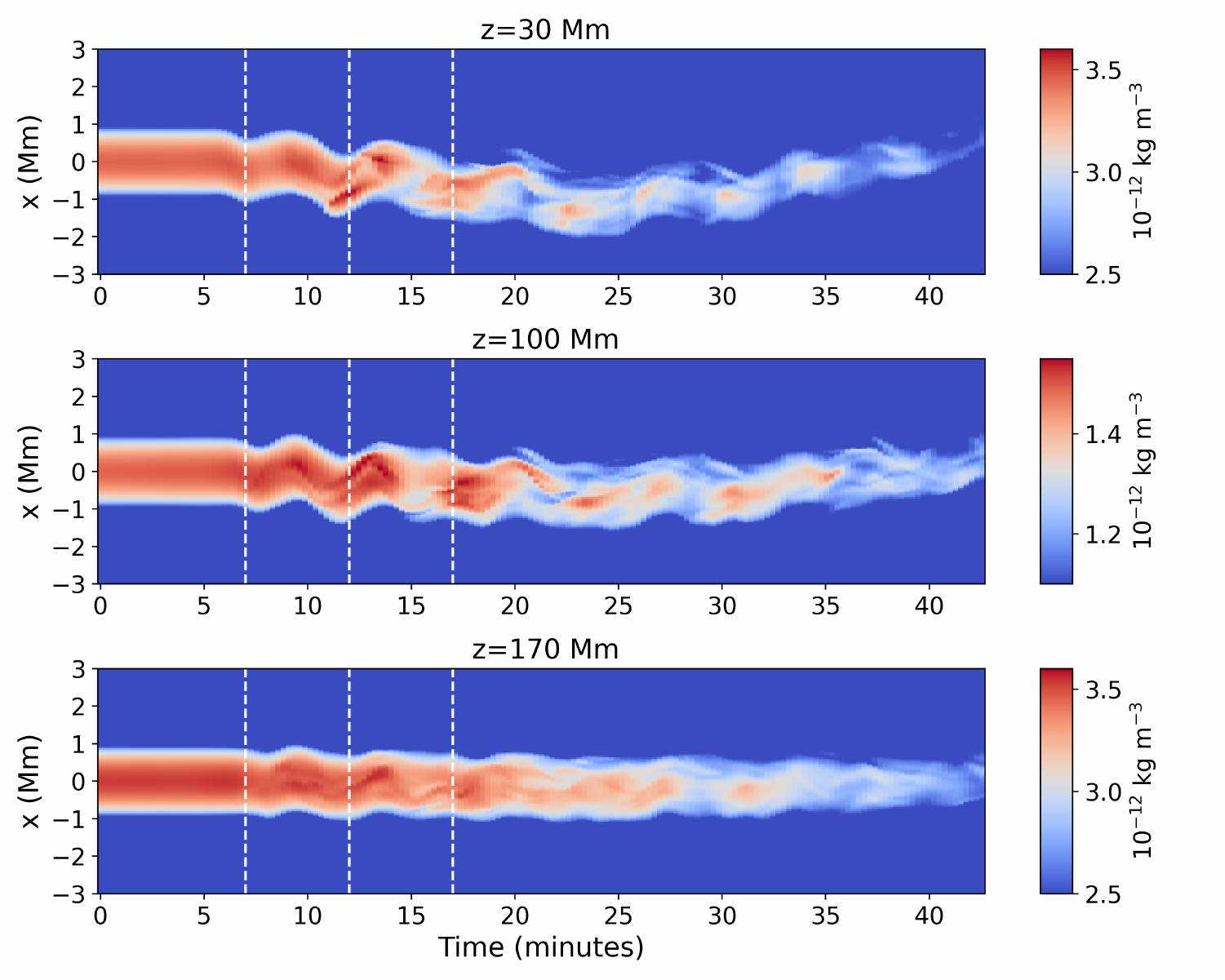}
    }
    \caption{Time distance maps of the density, integrated along the $y$-axis, for the our loop at different heights. The left panels are for the detrended driver and the right panels for the red noise driver. Vertical dashed lines are added at $t=7$, $12$, and $17$\,minutes.}
    \label{fig:timedist}
\end{figure*}

\begin{figure*}
    \centering
    \resizebox{\hsize}{!}{
    \includegraphics[trim={0.cm 1.5cm 0.4cm 0.4cm},clip,scale=0.55]{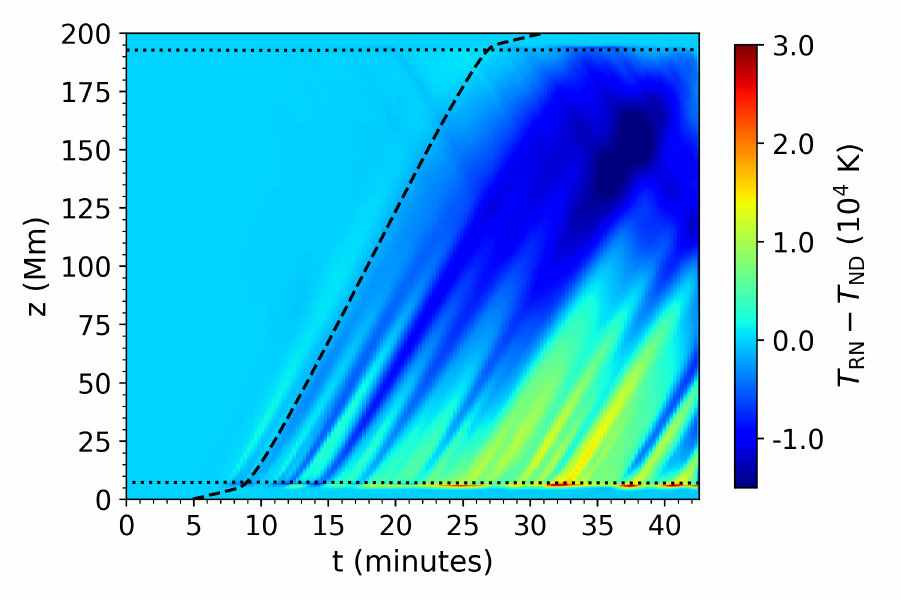}
    \includegraphics[trim={0.82cm 1.5cm 0.4cm 0.4cm},clip,scale=0.55]{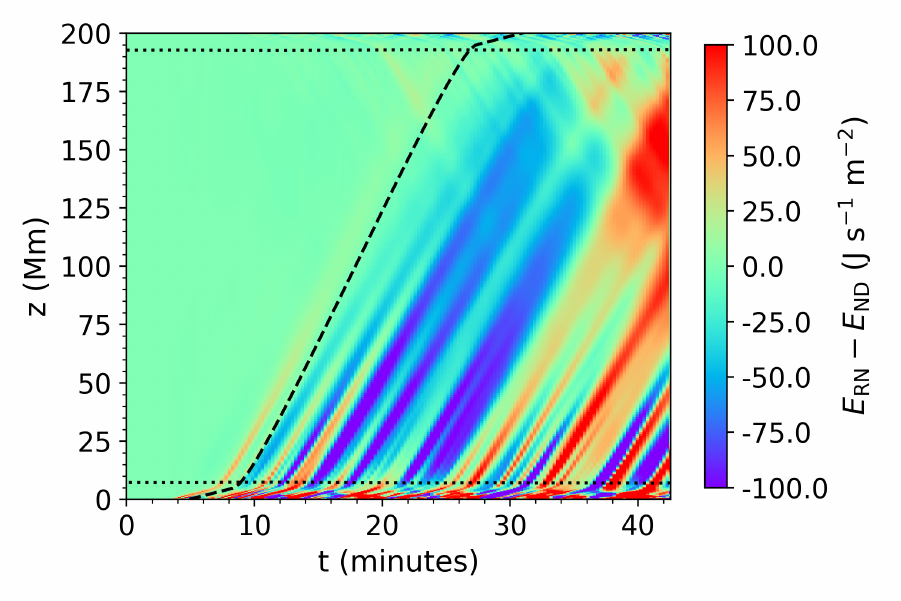}
    \includegraphics[trim={0.82cm 1.5cm 0.cm 0.4cm},clip,scale=0.55]{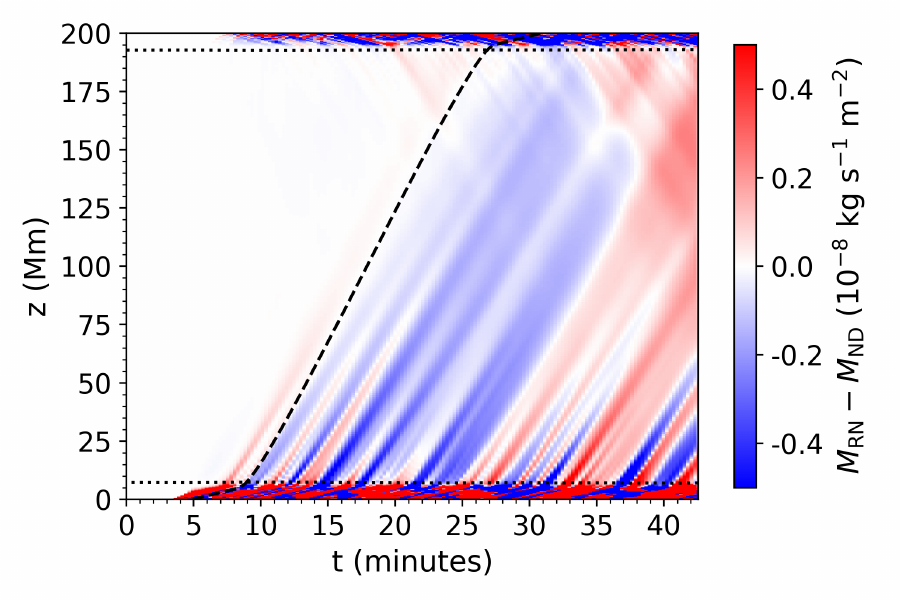}
    }
    \resizebox{\hsize}{!}{
    \includegraphics[trim={0.cm 0.4cm 0.4cm 0.4cm},clip,scale=0.55]{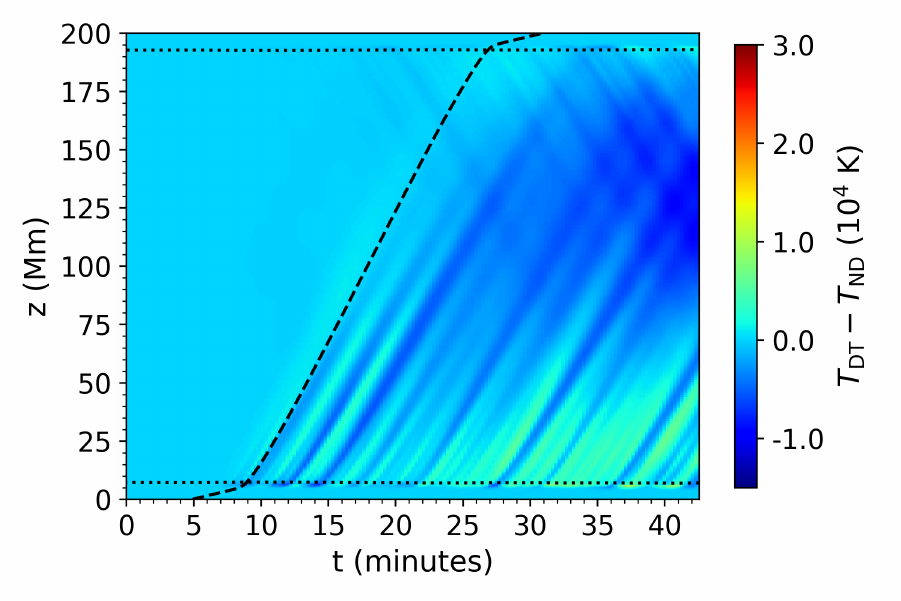}
    \includegraphics[trim={0.82cm 0.4cm 0.4cm 0.4cm},clip,scale=0.55]{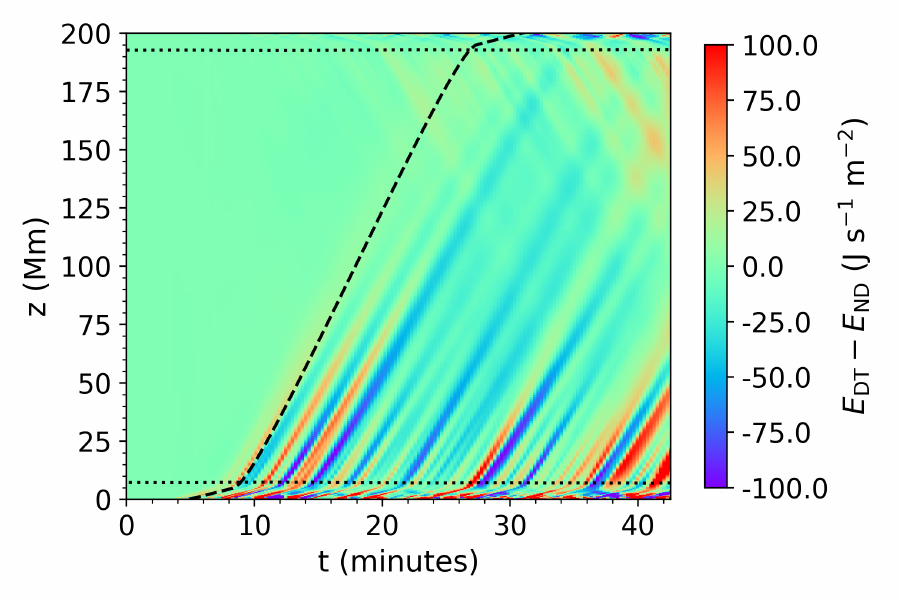}
    \includegraphics[trim={0.82cm 0.4cm 0.cm 0.4cm},clip,scale=0.55]{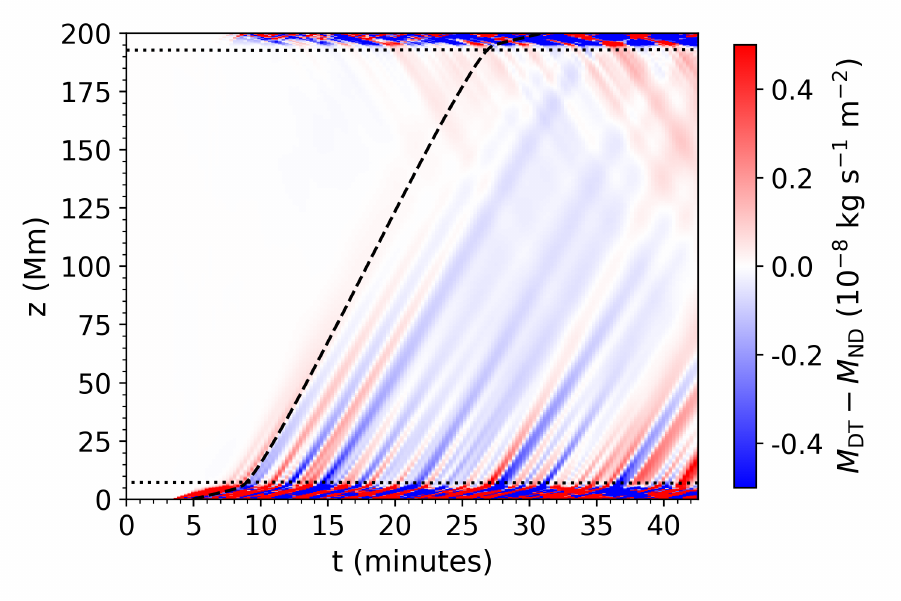}
    }
    \caption{Height vs time profiles of the temperature difference (left panels), enthalpy flux difference (middle panels), and mass flux difference (right panels), for the set-ups with the red noise driver ($\mathrm{RN}$, top) and detrended driver ($\mathrm{DT}$, bottom)  with respect to the non-driven case ($\mathrm{ND}$). The horizontal dotted black lines show the location of the $T=0.4$\,MK line. The black dashed lines show the estimated trajectory of a plasma parcel launched from one footpoint, with a varying velocity with height, equal to the average sound speed across the loop at each height, at $t=5$\,minutes.}
    \label{fig:temperature}
\end{figure*}

\begin{figure*}
    \centering
    \resizebox{\hsize}{!}{
    \includegraphics[trim={0.cm 0.cm 0.cm 0.cm},clip,scale=0.48]{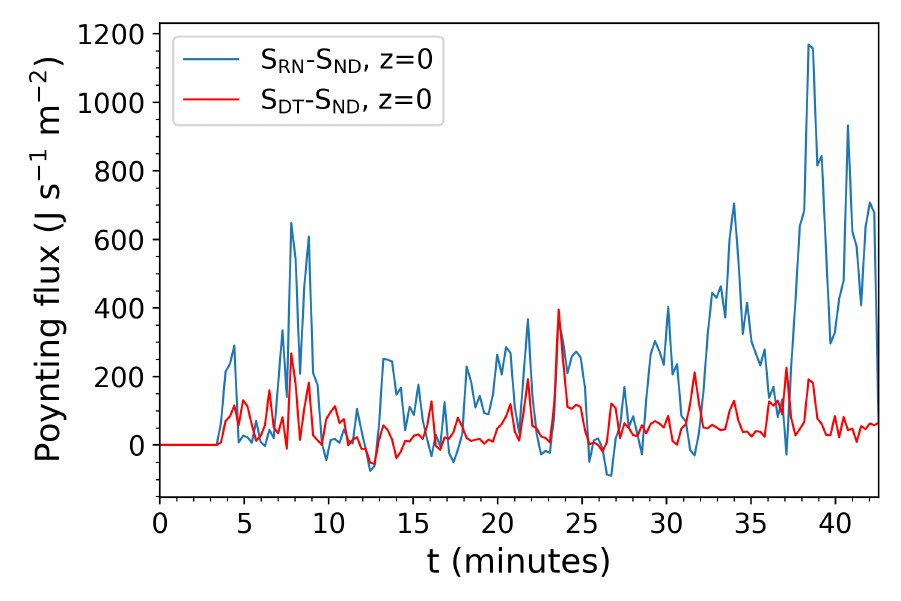}
    \includegraphics[trim={0.cm 0.cm 0.cm 0.cm},clip,scale=0.48]{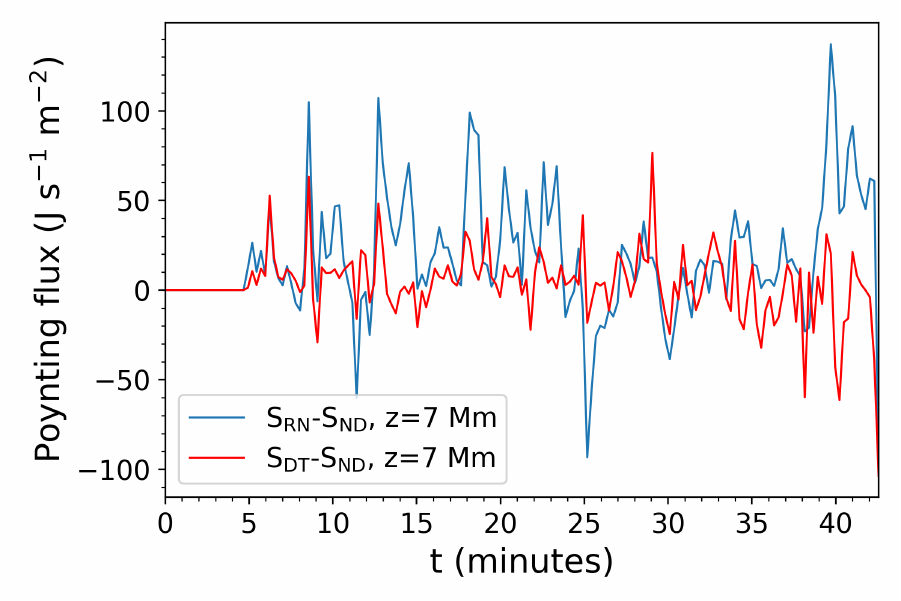}
    \includegraphics[trim={0.cm 0.cm 0.cm 0.cm},clip,scale=0.48]{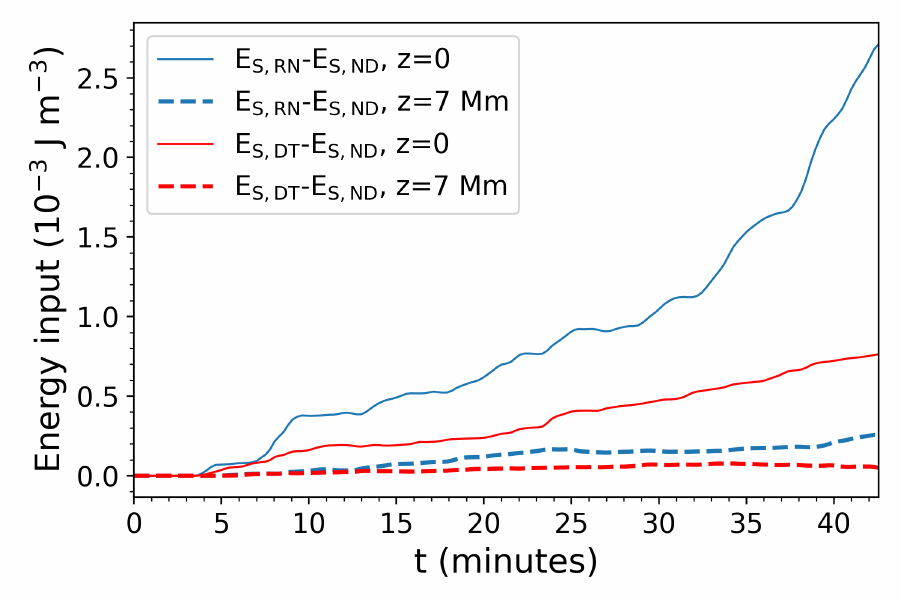}
    }
    \caption{The Poynting flux graphs (left and middle panels) and input energy density graphs (right panel) over time for the red noise driver (solid blue line, $\mathrm{RN}$ subscript) and the detrended driver (red line, $\mathrm{DT}$ subscript) relative to the non-driven case ($\mathrm{ND}$). The Poynting energy density input at the approximate height of the transition region ($z=7$\,Mm) is also given for both drivers (dashed lines).}
    \label{fig:energyinput}
\end{figure*}

\begin{figure}
    \centering
    \includegraphics[trim={0.cm 0.cm 0.cm 0.cm},clip,scale=0.5]{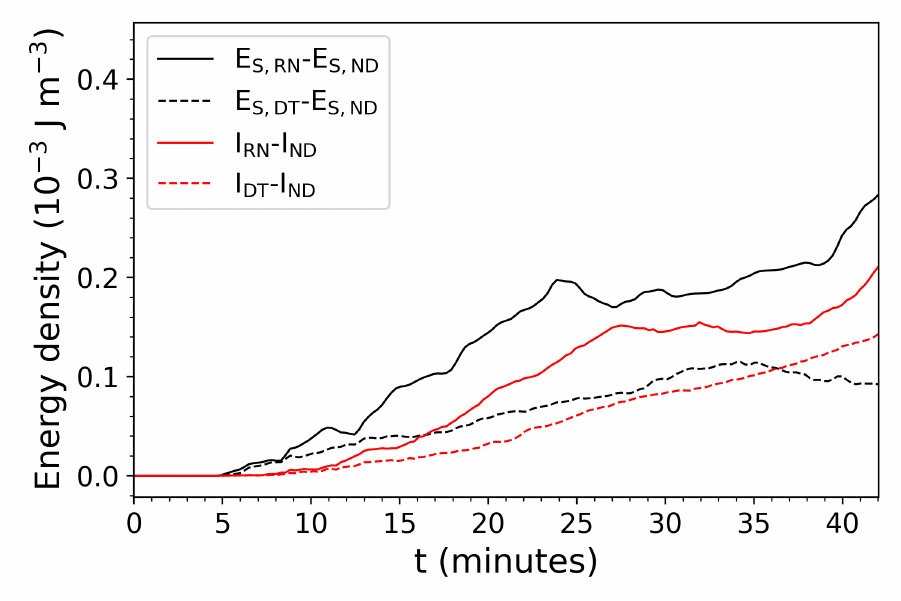}
    \caption{Energy density profiles over time, relative to the non-driven case  (ND). The solid lines correspond to the results for the red noise driver (RN)  and the dashed lines for the detrended driver (DT). The total Poynting energy input (black lines) from $z=7$\,Mm and $z=193$\,Mm and the internal energy (red lines) for the coronal part of the loop ($x\in[-6,6]$\,Mm, $y\in[0,4]$\,Mm and $z\in[7,193]$\,Mm) are shown.}
    \label{fig:energydensity}
\end{figure}

Our set-up consists of a 3D straight magnetic flux tube in a stratified atmosphere, modelling a coronal loop with its footpoints anchored in the chromosphere. Following the same methodology found in our past studies \citep{pelouze2023A&A...672A.105P,mingzhe2023ApJ...949L...1G,Karampelas2024A&A...681L...6K}, we first construct a 2.5D slice of our flux tube, in cylindrical coordinates, with initial conditions in hydrostatic equilibrium only along the vertical direction. We then let our set-up evolve and reach a quasi-equilibrium state, before interpolating it into a 3D cartesian grid.

We solve the 2.5D and 3D compressible MHD equations for a hydrogen plasma, using the PLUTO code \citep{mignonePLUTO2007}
\begin{align}
&\frac{\partial\rho}{\partial t} + \nabla \cdot (\rho \mathbf v) = 0\,,\\
&\rho \left[ \frac{\partial\mathbf v }{\partial t} + \rho(\mathbf v \cdot \nabla)\mathbf v \right] = - \nabla p + \rho\,g_z         + \frac{1}{\mu}\left(\nabla\times\mathbf{\mathbf B}\right) \times \mathbf B,\\
&\rho \left[ \frac{\partial \epsilon}{\partial t} + (\mathbf v \cdot \nabla)\epsilon \right] = -p \nabla \cdot \mathbf v + \eta \, \mu |\mathbf J|^2 + \nabla \cdot \mathbf{F}_c,\\
&\frac{\partial\mathbf B}{\partial t} = \nabla \times (\mathbf v \times \mathbf B) + \eta\nabla^2 \mathbf B,
\end{align}
where $\rho$, $p$, and $\mathbf v$ are the density, plasma pressure, and velocity, and $\mathbf B$ is the magnetic field. The electric current is $\mathbf J = \nabla\times\mathbf B /\mu$, with $\mu = 4 \pi \times 10^{-7}$\,H m$^{-1}$ being the magnetic permeability, $\eta$ the magnetic diffusivity, and $\epsilon = p/\left[\rho (\gamma-1)\right]$ the specific internal energy density, with $\gamma = 5/3$  the ratio of the specific heats. The quantity $\nabla \cdot \mathbf{F}_c$ represents the thermal conduction term. Finally, $g_z$ is the gravitational acceleration along the z direction. No additional source terms such as radiative cooling or background heating are included in our models.

We satisfy the solenoidal constraint for the magnetic field ($\nabla\cdot\mathbf B=0 $) through a hyperbolic divergence cleaning method, with the extended generalized Lagrange multiplier (GLM) formulation. We use the third-order Runge-Kutta method to calculate the time step, and the linearised Roe Riemann solver for computing the fluxes. For the spatial reconstruction method, we use the second-order finite volume piecewise parabolic method (PPM) for the 2.5D set-up, and the fifth-order monotonicity preserving scheme (MP5) with a second-order global accuracy for the 3D set-up. Only in the 2.5D set-up did we add explicit magnetic diffusivity to improve the stability of the code, measured in code units as $\eta = R_m^{-1} = 10^{-4}$, where $R_m$ is the magnetic Reynolds number. The finite size of our grid in both the 2.5D and 3D cases also gives rise to effective numerical diffusivity, estimated of the order of $10^{-5}$ to $10^{-4}$ through a parameter study, as well as our past findings \citep[e.g][]{karampelas2019}. 

We include thermal conduction, calculating the value for the parallel thermal conduction coefficient (in J\,s$^{-1}$\,K$^{-1}$\,m$^{-1}$) from the Spitzer conductivity \citep{Orlando2008ApJ}, with the method developed by \citet{LinkerEtAl2001, LionelloEtAl2009} and \citet{MikicEtAl2013},
\begin{equation}    
\kappa_{\parallel} = 9 \times 10^{-12} \left \{
    \begin{array}{ll}
    T^{5/2}, & \text{if $T>T_\mathrm{cut}=0.25$\,MK}, \\
    T_\mathrm{cut}^{5/2}, & \text{if $T\leq T_\mathrm{cut}=0.25$\,MK}, 
    \end{array}\label{eq:kpar}\right.
\end{equation}
where we consider a fixed cutoff temperature $T_\mathrm{cut}=2.5\times 10^5$\,K, which is a value commonly used in past studies \citep[e.g.][]{JohnstonBradshaw2019,pelouze2023A&A...672A.105P}. The above treatment for $\kappa_{\parallel}$ leads to an artificial broadening of the transition region as the system evolves. As discussed in \citet{pelouze2023A&A...672A.105P}, this saves us the computational costs of resolving the very steep temperature gradient of the regular transition region between the solar chromosphere and the corona, which would normally require a prohibitively high resolution along the height ($\sim 1$\,km). The broadened transition region has a minimum temperature scale length of $1.6$\,Mm \citep{JohnstonBradshaw2019}, which can easily be resolved by 2D and 3D simulations. Finally, the PLUTO code accounts for saturation effects when very large temperature gradients are considered. The conductive flux $\mathbf{F_c}$ varies in the code between the classical and saturated regimes ($\mathbf{F}_\mathrm{cl}$ and $F_\mathrm{s}$),
\begin{equation}
\mathbf{F_c} = \frac{F_\mathrm{s}}{F_\mathrm{s}+|\mathbf{F}_\mathrm{cl}|}\,\mathbf{F}_\mathrm{cl},\, \mathbf{F}_\mathrm{cl} = \kappa_{\parallel} \hat{\mathbf{b}}\left( \hat{\mathbf{b}} \cdot \nabla T \right), \,
F_\mathrm{s} = 5\rho\phi\,c_\mathrm{iso}^3,
\end{equation}
where $\hat{\mathbf{b}}$ is the unit vector along the magnetic field, $c_\mathrm{iso}$ is the isothermal sound speed, and $\phi$ is a free parameter with a default value of $0.3$ \citep[see][for a detailed description]{mignonePLUTO2007,mignonePLUTO2012}.

\subsection{2.5D flux tube: Initial and boundary conditions}\label{sec:initial2D}
\textit{Initial conditions:} We consider a 2.5D domain in cylindrical coordinates $(r,z)$, with size of $r\in [0,8]$\,Mm and $z\in [0,200]$\,Mm and a uniform grid of $200\times 2048$ cells. To construct our flux tube, we consider an initial straight magnetic field $B_z=30$\,G and we use a temperature profile, derived from \citet{AschwandenSchrijver2002},
\begin{align}
&T(r,z) = T_\mathrm{Ch} + (T_\mathrm{C}(r) - T_\mathrm{Ch})( 1 - \left[ (L-z)/(L-\Delta_\mathrm{Ch}) \right]^2 )^{0.3},\\
&T_\mathrm{C}(r) = T_\mathrm{C,e} + (T_\mathrm{C,i} - T_\mathrm{C,e})\, \zeta(r),
\end{align}
for height $\Delta_\mathrm{Ch} \leq z \leq 200-\Delta_\mathrm{Ch}$, where $L=200$\,Mm is the loop length and $\Delta_\mathrm{Ch}=5$\,Mm is the width of our chromosphere; $T_\mathrm{Ch} = 0.02$\,MK is the temperature of the chromospheric part of the loop and $T_\mathrm{C}(r)$ is the temperature profile at the apex ($z=100$\,Mm); and $T_\mathrm{C,i} = 1$\,MK and $T_\mathrm{C,e} = 1.5$\,MK are the temperature values inside and outside of the flux tube at the apex, respectively. The function $\zeta(r)$ gives us the flux tube profile along the $r$ direction, with $R=1$\,Mm being the radius of the flux tube cross-section:
\begin{equation}
\zeta(r) =0.5\left[ 1 - \tanh\left(\left( \left[r/R\right]-1 \right) 20 \right) \right].
\end{equation}
The density profile at the bottom of the chromosphere ($z=0$ and $z=200$\,Mm) is calculated as
\begin{equation}
\rho_\mathrm{Ch} = \rho_\mathrm{Ch,e} + (\rho_\mathrm{Ch,i}-\rho_\mathrm{Ch,e})\, \zeta(r),
\end{equation}
where $\rho_\mathrm{Ch,i} = 3.51 \times 10^{-8}$\,kg m$^{-3}$ and $\rho_\mathrm{Ch,e} = 1.17 \times 10^{-8}$\,kg m$^{-3}$ are the density values inside and outside the footpoint. Finally, we consider sinusoidal gravity ($g_z(z) = 274\, \cos(\pi\,z/200)$ in m s$^{-2}$) along the $z$-axis, to account for the change in gravity projected along a semi-circular coronal loop. We solved the equations of the hydrostatic equilibrium along the $z$ direction using a  forward-Euler method. To avoid asymmetries in the z-direction in the initial conditions, we took into account the expected symmetry of our model with respect to the z=100 Mm plane at the apex. In the left panel of Figure \ref{fig:hydrostatic}, we show the density, temperature, and magnetic field profiles inside and outside of the flux tube, up to $20$\,Mm, with the transition region shown at $5$\,Mm.

\textit{Boundary conditions:} We consider axisymmetry at $r=0$ and open boundary conditions at $r=8$\,Mm for all quantities. At the bottom ($z=0$) and top ($z=200$\,Mm) boundaries, we take the zero-gradient condition for the three magnetic field components ($B_r, B_{\theta}, B_z$) and consider antisymmetric conditions for the three velocity components. We also consider a constant temperature $T_\mathrm{Ch}$ and symmetric boundary conditions for the density. This effectively prevents any flows across the top and bottom boundary from taking place.  

\textit{2.5D MHD relaxation:} Since our initial 2.5D set-up was not in hydrostatic equilibrium in the $r$ direction, we let it evolve and reach a semi-equilibrium state for a total time of $t=3890$\,s. Between $t=778$\,s and $t=3112$\,s, we artificially reduced the value of the velocity components per iteration over the entire domain, by using $v_i = v_i/n_d$, with $n_d=1.0001$ and $i=(r,\theta,z)$. The vertical profiles of the magnetic field, density, and temperature after the relaxation process are seen in the right panel of Figure \ref{fig:hydrostatic}, where     the widening of the transition region can also be seen. The residual values for the velocities in our set-up have a maximum value of $\sim 4$\,km s$^{-1}$ along the z direction and and $\sim 0.05$\,km s$^{-1}$ along the x and y directions. These values are higher than those reported in past studies \citep[see][]{pelouze2023A&A...672A.105P}, where the resulting values where $0.5$ km s$^{-1}$ after $\sim 47000$\,s of relaxation. However, our set-up has different boundary conditions and our method allows for a narrower transition region \citep[similarly to][]{mingzhe2023ApJ...949L...1G} and for a temperature, density, and magnetic field profile closer to those initially assigned. In addition, once we implemented a footpoint driver into our set-ups, the generated velocities far exceeded these residual velocities. During the relaxation, we have restructuring of the magnetic field both via advection and through dissipation since we have non-zero resistivity present. The magnetic field is no longer purely vertical, but now consists of a radial ($B_r(r,z)$) and a vertical ($B_z(r,z)$) component. The radial component has its maximum values near the footpoints, which are an order of magnitude lower than values of the vertical component there. The relaxation also leads to a radial variation of the density and temperature profiles. As an example, Figure \ref{fig:DTprofile} shows the radial density and $B_z$ magnetic field profiles at different heights at the end of the 2.5D relaxation.

\subsection{3D  flux tube: Initial and boundary conditions}\label{sec:initial3D}
\textit{Initial conditions:} We create a straight flux tube in cartesian coordinates $(x,y,z)$, by interpolating the post relaxation state of our 2.5D slice into the 3D grid. Our new domain has a size of $x\in [-6,6]$\,Mm, $y\in [0,4]$\,Mm and $z\in [0,200]$\,Mm with uniform spacing in the $x$ and $y$ directions and $\delta x = \delta y = 0.040$\,Mm. Along the $z$ direction we have uniform grids with $\delta z = 0.098$\,Mm for $z\leq 10$\,Mm and $z\geq 190$\,Mm and another uniform grid with $\delta z = 0.8$\,Mm for $24\,\text{Mm}\leq z\geq 176$\,Mm. Finally, we consider stretched  grids of $40$ grid points each for $10\,\text{Mm}\leq z\geq 24$\,Mm and $176\,\text{Mm}\leq z\geq 190$\,Mm. This provides us with the resolution, up to a height of $\sim 15$\,Mm, to treat the transition region, while further reducing the computational costs. We  note here, that in our simulations we never have plasma with a transition region temperature and density above $z=15$\,Mm.

\textit{Boundary conditions:} At both boundaries in the $x$ direction, and at the $y=4$\,Mm boundary we employ open boundary conditions for all quantities, while we are using reflective boundary conditions at the $y=0$ boundary. This enforces a symmetry to our system, which allows us to simulate only half the flux tube, given our driver of choice. For the top ($z=200$\,Mm) boundary, we use the same conditions as in the 2.5D case. For the bottom ($z=0$) boundary, we use the same conditions until $t=202$\,s, allowing our system to settle to the 3D configuration and avoid any numerical issues that could arise. 

\textit{Driver:} After $t = 202$\,s, we change the conditions for $v_x$ and $v_y$ by applying a broadband, linearly polarised driver:
\begin{align}
&\lbrace v_x,v_y\rbrace = \lbrace V(t)\zeta(r,t),0\rbrace
,\\
&\zeta(r) =0.5\left[ 1 - \tanh\left(\left( \left[r/R_\mathrm{d}\right]-1 \right) 20 \right) \right].
\end{align}
Here, in function $\zeta(r,t)$, we consider $R_\mathrm{d}=2.5$\,Mm as the radius of the driver (i.e. the distance from the tube axis where $v_x\neq 0$, which  quickly drops to zero past that radial distance). For radii larger than $R_d$, $v_x$ quickly drops to zero. The time dependence of $\zeta(r,t)$ comes from our driver tracking the tube axis, while the latter is moving. The location of the driver is controlled by
\begin{equation}
r(t) =\sqrt{\left( x - x_0(t) \right)^2 + y^2},
\end{equation}
where $x_0(t)$ (and $y_0(t)=0$) is the centre of each driver, calculated by numerically integrating the respective velocity signal ($V(t)$) over time. For the velocity signal we use the Python package \textit{colorednoise 2.1.0} to create a red noise signal with a power spectral density $S\propto f^{-1.66}$, where $f$ is the frequency \citep[see also][]{afanasyev2020decayless}. This results in a velocity profile that can be viewed practically as random motions along the $x$ direction. The use of drivers with power-law spectra of red noise type  is inspired by observations of sunspot oscillations and the dynamics of magnetic bright points \citep{Kolotkov2016A&A...592A.153K,Abramenko2011ApJ...743..133A,Chitta2012ApJ...752...48C}. The latter, as suggested in Afansyev et al. 2020, can be regarded as the motion of loop footpoints. The driving signal $V(t)$ is shown for two different cases in Figure \ref{fig:Vprofile}. In blue we show the original red noise velocity signal and spectrum. The dashed black line shows the background trend of the red noise signal, calculated with a low pass Gaussian filter. By subtracting this background trend from the red noise signal, we get the detrended signal, which is plotted in orange. The RMS velocity ($V_{RMS}$) of the detrended signal ($\sim 0.94$\,km s$^{-1}$) are comparable to the RMS velocities of horizontal motions of magnetic bright points ($\sim 1.32$\,km s$^{-1}$) derived from SST and \textit{Hinode} observations \citep{Chitta2012ApJ...752...48C}. For the red noise signal, the peak velocity amplitudes are comparable to those from wave drivers in \citet{Howson2022A&A...661A.144H}. The spectrum of the detrended signal shows a maximum power between $\sim 2$ to $3$\,mHz and has reduced power at the lower frequencies, with respect to the initial signal. Hereafter,  we   denote the driver with the original signal   the  red noise  driver and the other  the detrended driver. We also run a simulation without a velocity driver, to isolate the evolution of the system due to the lack of a perfect initial equilibrium.

\section{Results} \label{sec:results}
After we interpolated our 2.5D slice onto the 3D cartesian grid, we then switched on the footpoint driver at $t> 202$\,s, and until $t_{max}= 2552$\,s or $42.5$\,minutes. In our analysis presented below, we   focus  exclusively on the coronal part of our set-up. Because we lack  a more realistic profile, the chromospheric part of our loop  is used as a mass reservoir of quasi-constant temperature, and is used to anchor our loop footpoints.

Figure \ref{fig:Comp} shows the integrated density and temperature along the y-axis, as well as the $v_z$ velocity profile on the $y=0$ plane at the end of the each simulation, for the set-ups with the detrended driver (top panels) and the red noise driver (bottom panels). In the animations, an additional drop in the temperature of the coronal plasma is observed, mainly outside of the loop. The latter is the combined effect of the thermal conduction along the nearly vertical magnetic field, and the initial non-zero $v_z$ velocity profiles of our system since our set-up is not in perfect pressure and thermal equilibrium \citep[see also the models by][]{pelouze2023A&A...672A.105P,mingzhe2023ApJ...949L...1G}, and  is also the reason why we  observe a drop in the values of the density over time. Looking at the mean density profiles over time for different coronal heights in Figure \ref{fig:Density}, we see this drop as manifesting in the non-driven case (i.e. for our loop without a driver). The observed drop in the average density is of the order of $\sim 13\%$ for $z\in[20,\,180]$\,Mm. The three different heights chosen are $z=100$\,Mm (apex), $z=20$\,Mm, and $z=180$\, Mm. The latter two showcase the symmetry of the non-driven case with respect to the apex.

\subsection{Wave excitation and propagation}
Using the previously described broadband drivers, we excite linearly polarised waves of different frequencies into our domain. \citet{pelouze2023A&A...672A.105P} have shown that the cutoff of transverse waves at frequencies above $2$\,mHz only results in a weak attenuation, with the waves being able to transport energy from the chromosphere to the corona. Given the range of frequencies for the two drivers, both of which exhibit similar power distribution between $2$\,mHz and $3$\,mHz, we expect the drivers to excite propagating transverse waves that will reach the coronal part of the loop, forming a standing oscillation pattern.

From the accompanying animations of Figure \ref{fig:Comp}, we see that the drivers generate a large number of transverse and longitudinal waves that superimpose and set up oscillations in each loop. The first thing we note is the existence of localised perturbations in the temperature, density (less visible), and $v_z$ profiles that can be described as longitudinal wave pulses. Due to the speed of their propagation along the magnetic field, these perturbations are expected to be slow waves, as we show later on via our analysis.

Focusing on the transverse waves, the accompanying animations of Figure \ref{fig:Comp} for the case with the detrended driver reveal an apparent standing transverse oscillation for our loop. Similarly, transverse oscillations are shown to be generated by the red noise driver. The power distribution of the latter, with its increased contribution from the lower frequencies, leads to a transverse displacement of the loop footpoint at $z=0$, in addition to what again appears to be a standing transverse oscillation. In \citet{Karampelas2024A&A...681L...6K}, transverse standing waves described as decayless oscillations were reportedly generated in simulation by the same red noise driver for various loop set-ups. Figure \ref{fig:Vxapex} shows the $v_x$ velocity profiles calculated from the oscillating signals of the centre of mass displacement for the two loops, generated by the two drivers. We see that  these two signals exhibit a very similar behaviour, with comparable velocity peak amplitudes, as well as RMS velocities of $\sim 2.5$\,km s$^{-1}$ and $\sim 2.3$\,km s$^{-1}$ for the loop driven by the red noise and detrended driver, respectively. The frequencies of the oscillations   match those of the fundamental standing kink mode and its harmonics, with Fourier power spectral density (PSD) profiles for the centre of mass displacement similar to those shown in Figure \ref{fig:Fourier}. The left profile shows the PSD of the original centre of mass displacement and the middle profile shows the PSD of the detrended displacement, after applying a high pass filter to remove the low-frequency motions. We can see that the spectrum of the detrended signal from the centre of mass displacement, excited by the red noise driver, is qualitatively very similar to the spectrum of the loop oscillation excited by the detrended driver.

\subsection{Development of the Kelvin-Helmholtz instability}
The top panel of Figure \ref{fig:KHI} shows the last snapshot of the simulation ($t=2554$\,s) for our oscillating loop for the detrended driver. On the left side of that panel, we see the integrated density along the $y$ direction for the coronal loop. Also visible is the contour line for $T=0.4$\,MK for the integrated temperature along the y direction, which we use as an (extreme) upper limit for the broadened transition region in our set-up. On the right side of that panel, we see the loop cross-section for the density at different coronal heights. In the accompanying animation, we see that the loop cross-sections shown here are fully deformed by the development of the KH instability, all along the coronal part of the loop. Thus, we show that our broadband kink wave driver leads to the manifestation of the KH instability across the entire volume of the coronal part of the loop, leading to fully deformed 3D loops. In the bottom panel of Figure \ref{fig:KHI}, we also see the last snapshot for an oscillating loop driven by a red noise-type  driver. Again, we report the development of the KH instability and the complete deformation of the loop cross-section along the coronal part of the loop, alongside the aperiodic footpoint displacement that exceeds the amplitude of the loop oscillation. 

The temporal evolution of the KH instability is shown for both drivers in Figure \ref{fig:timedist}, through time-distance maps ($t-x$) of density integrated along $y$. The left panels deal with the loop response to the detrended driver, and the right panels with the loop response to the red noise driver. We  placed slits along $x$, at three different heights for our domain ($z=30,\,100$, and $170$\,Mm). We observe the gradual drop in the average density of the loop apex over time. This is mainly due to the mixing of the loop plasma with the hotter and less dense background coronal plasma (from the KH instability), and to a lesser extent due to the ponderomotive force associated with slow waves generated by the driver and the effects of the thermal conduction to the energy balance of the system \citep{mingzhe2023ApJ...949L...1G}. We see the displacement of the driven footpoint for the red noise driver, caused by the low-frequency motions of the driver.  We also detect a small initial phase difference in the displacement among the driven footpoint, the loop apex, and the anchored footpoint, due to the waves propagating along the loop from the driven footpoint. This difference  effectively vanishes after $1.5$ periods (or at $t\sim 17$\,minutes), due to the formation of the standing wave.

Signs of the KH instability are present for both drivers; they are in fact slightly more prominent for the loop with the red noise driver, thanks to the additional driver energy causing stronger shear velocities. However, both simulations exhibit a qualitatively similar behaviour in the spatial and temporal evolution of the KH instability. Looking at the apex ($z=100$\,Mm), we see that the instability starts developing properly after $\sim 1-1.5$ wave periods, which is after the standing wave has formed. It is interesting to note here that some fainter traces of density maxima, which seem to be caused by the KH instability, appear by the time the propagating transverse kink wave reaches the apex.. This can also be seen in either animation of Figure \ref{fig:KHI}. From the linear analysis, we do not expect the propagating waves to be KH unstable \citep[][]{heyvaerts1983}. However, the strong velocity shear from the  unidirectional propagating kink waves, which is also amplified by resonant absorption \citep[e.g.][]{antolintvd2019FrP.....7...85A}, together with the density structure transverse to the mean magnetic field, could non-linearly lead to a weak manifestation of the KH instability, similarly to and alongside the manifestation of uniturbulence there \citep[e.g][]{magyar2017,tvd2020ApJ...899..100V}.

Both simulations exhibit a small delay to the manifestation and development of the KH instability near the anchored footpoint, with respect to the loop apex. This delay, which is smaller than $1$\,min, can be identified as a small phase difference in the appearance of the density maxima in between the $z=100$\,Mm and $z=170$\,Mm panels in Figure \ref{fig:timedist}, during the first two periods of oscillation. We also see an apparent delay in the manifestation of the KH instability at the driven footpoint with respect to the anchored footpoint. However, examining the animations of Figure \ref{fig:KHI} reveals no such delay; instead, these apparent effects are due to compression near the driven footpoint, saturating the signal in the first two periods.

\subsection{Energy evolution of the oscillating loops}
In order to study the effects of our drivers on the energy evolution of the system, we calculate the surface average temperature ($T_i$), enthalpy flux ($E$), and mass flux ($M$) for our set-ups, at each height for the duration of the simulations:
\begin{align}
    &T_i(z,t) = \frac{1}{A}\int_{A'} T_i(z,t) dA', \\
    &E_i(z,t) = \frac{1}{A}\int_{A'} \frac{\gamma}{\gamma-1}p_iv_{z,1} dA', \\
    &M_i(z,t) = \frac{1}{A}\int_{A'} \rho_i v_{z,i} dA'.
\end{align}
Here $i$ corresponds to the red noise driver ($\mathrm{RN}$), the detrended driver ($\mathrm{DT}$), and the non-driven case ($\mathrm{ND}$); $\gamma = 5/3$; and $A$ is the surface area of the $xy$ plane. Figure \ref{fig:temperature} shows the difference in the mean temperature ($T_\mathrm{RN,DT}-T_\mathrm{ND}$), enthalpy flux ($E_\mathrm{RN,DT}-E_\mathrm{ND}$), and mass flux ($M_\mathrm{RN,DT}-M_\mathrm{ND}$) for both driven set-ups ($\mathrm{RN}$, top panels; $\mathrm{DT}$, bottom panels), with respect to the non-driven case ($\mathrm{ND}$). The dotted black lines trace the height where $T=0.4$\,MK, which we consider as the uppermost limit of the transition region. 

In these  panels we detect perturbations generated and propagating from the driven footpoint, which are also reflecting at the anchored footpoint, similarly to past cases \citep[e.g.][]{mingzhe2019,mingzhe2023ApJ...949L...1G,karampelas2019}. The dashed black lines in the panels show the calculated trajectory of a plasma parcel along the z direction, when we consider a varying velocity with height, corresponding to the profile along the z-axis of the average sound speed at $t=5$\,minutes. These perturbations are almost parallel to the dashed lines. This shows that these perturbations, which were also detected in Figure \ref{fig:Comp}, are in fact slow waves that propagate along the near-vertical magnetic field. 

From the difference of the mean temperature, in the left panels of Figure \ref{fig:temperature}, we see a drop in the mean temperature along the loop over time, which is more prominent for the set-up with the red noise driver. This drop in the mean temperature is expected in part due to thermal conduction and the thermal imbalance between the hotter coronal and the colder chromospheric parts, as mentioned at the beginning of Section \ref{sec:results}. The development of the KH instability   also leads to an apparent drop in the average temperature, due to the mixing of the cooler loop plasma with the hotter background coronal plasma, as expected from our past studies of colder loops embedded in a hotter corona \citep[e.g.][]{karampelas2017}. In the same panels, we also see a symmetry along the z direction in the variation of the relative  temperature profile that suggests an adiabatic response of our systems to the effects of the ponderomotive force along the oscillating loop, which has also been observed in past studies \citep{terradasofman2004ApJ,karampelas2019,VanDamme2020A&A...635A.174V,mingzhe2023ApJ...949L...1G}. These effects are stronger for the case of the red noise driver, due to the stronger perturbations that it imposes to the driven footpoint. The detected gradual increase in the temperature at the transition region, above the driven footpoint, is also associated with the launch of slow waves by that propagate along the loop, dissipating their energy. This can be seen in Figure \ref{fig:Comp}, where we relate the increase in temperature and density to the slow waves shown in the $v_z$ profile. 

The middle and right panels of Figure \ref{fig:temperature} show  similar behaviour for the enthalpy and mass flux along the loop. We see an initial enthalpy and mass flux towards the footpoints, due to the ponderomotive force, alongside small spikes of enthalpy and mass flux from the driven footpoints to the apex. These spikes, which follow the trajectory of the slow waves, are related to the respective temperature increase from the dissipation of the slow waves, in our model. The stronger enthalpy and mass fluxes are connected to higher values of temperature increase at the driven footpoint. As a final note, we  saturated the colourmaps in the plots for the mass flux, in order to reveal the evolution of the coronal part of the loop. The density in the transition region and chromosphere at the loop footpoints is orders of magnitudes higher than in the coronal part and as a result  exhibits higher values for the  mass flux.

We  also calculated the surface averaged Poynting flux ($S$) and the cumulative electromagnetic energy density input due to the Poynting flux ($E_S$) from each driver over time:
\begin{align}  
    &S(t) = - \frac{1}{A}\int_{A_1'} \frac{1}{\mu_0}[(\mathbf{v}\times \mathbf{B})\times \mathbf{B}]\cdot d\mathbf{A_1'},\\ 
    &E_S(t) = - \frac{1}{V}\int^t_{0}\int_{A_1'} \frac{1}{\mu_0}[(\mathbf{v}\times \mathbf{B})\times \mathbf{B}]\cdot d\mathbf{A_1'} dt'.
\end{align}
Here $A_1$ is the surface area of the xy-plane (bottom and/or top boundary), $V$ is the volume of our domain, and $d \mathbf{A_1'}$ is the normal vector to the xy-plane. In the left panel of Figure \ref{fig:energyinput} we plot the Poynting flux over time at the bottom boundary ($z=0$) for the red noise driver (with blue) and the detrended driver (with red), after subtracting the Poynting flux generated by the residual non-zero velocities of the non-driven case. In the middle panel of the same figure, we also plot the respective Poynting fluxes at the $z=7$\,Mm, which is the average height of the top of our broadened transition region (i.e. the base of the coronal part of the simulation domain) at the driven footpoint. We see that the red noise driver generates a stronger Poynting energy flux to our system than the detrended driver, which leads to a higher total Poynting energy density input into our system, from the $z=0$ plane, as seen in the right panel of Figure \ref{fig:energyinput}. In the same panel we also depict the cumulative Poynting energy density input at the approximate position of the base of the corona ($z=7$\,Mm). We see that only a small part of the total input energy passes into the corona in the form of a Poynting flux. The reduction of Poynting flux between $z=0$ and $z=7$\,Mm implies that energy has been deposited between the driving boundary and the base of the corona, which is expected to lead to plasma heating and evaporation in the chromosphere and the transition region. Such  behaviour is already detected in Figure \ref{fig:temperature}, as a combination of wave heating and the adiabatic effects from the ponderomotive force, and is more prominent in the case of the red noise driver, which explains the stronger enthalpy and mass fluxes along the loop.  Additional physics, such as radiation and partial ionisation effects, are required to better quantify the chromospheric response to the heating. However, our set-up shows that a significant part of wave energy dissipation will take place in the lower solar atmosphere, and its response in the evolution of the coronal energetics should not be ignored, as is often the case in theoretical studies of purely coronal loops.

Finally, to better understand the energy evolution of our systems, we calculate the internal energy density in the coronal part of the loop ($x\in[-6,6]$\,Mm, $y\in[0,4]$\,Mm and $z\in[7,193]$\,Mm):
\begin{align}
    &I(t)=\frac{1}{V}\int_{V'} \frac{p}{\gamma -1}dV' - F(t),\\
    &F(t) = \frac{1}{V}\int^t_{0}\int_{A'}\left( \frac{\rho v^2}{2}+\rho \Phi + \frac{\gamma p}{\gamma -1} \right)\mathbf{v}\cdot d\mathbf{A'}dt.
\end{align}
Here $F(t)$ is the energy flux through the boundaries due to the plasma flow, $V$ is the volume of the domain, $\Phi$ is the gravitational potential, and $d\mathbf{A'}$ is the normal vector to each boundary. Following the same methodology as in past studies \citep[e.g.][]{karampelas2019,mingzhe2023ApJ...949L...1G}, we subtract the energy flux through every boundary from the internal energy density in order to compensate for energy variations due to the plasma and enthalpy flux in our system, and focus on the energy variation due to the driving. In Figure \ref{fig:energydensity} we plot the compensated internal energy ($I$) for the coronal part of the loop (red lines) for the two different cases (with the red noise driver $\mathrm{RN}$ and the detrended driver $\mathrm{DT}$) relative to the non-driven case ($\mathrm{ND}$). Alongside them, we plot the cumulative Poynting energy input ($E_S$). Unlike in Figure \ref{fig:energyinput}, here we calculate this input at $z=7$\,Mm and at $z=193$\,Mm, taking into account the response of the non-driven footpoint to the oscillations. The profiles  in Figure \ref{fig:energydensity} show a stronger increase in the relative internal energy of the loop, when the red noise driver is considered. Since the red noise driver results in more energy input into the corona than the detrended driver, this higher increase indicates that wave energy dissipation is at least in part responsible for this apparent heating that we detect. In the case of the detrended driver, we detect a drop in  the Poynting input that is not followed by a decrease in the compensated internal energy. Although this seems peculiar at first glance, it can be explained when considering two things. First of all, the input energy is introduced in the form of waves, the dissipation of which contributes to the increase in the internal energy. Given the observed delayed response of the internal energy profiles to the Poynting input, as seen for both cases, a temporary drop in the cumulative Poynting energy input due to an increase in negative Poynting fluxes will not manifest directly in the internal energy evolution. The second thing  to note is that our loop exhibits higher internal energy density values than the surrounding plasma, as can be calculated from the initial conditions. Therefore, the KH instability-induced mixing of plasma and the plasma flows along the loop will lead not only to the apparent drop in temperature discussed in Figure \ref{fig:temperature}, but also to an apparent increase in the average internal energy density of our system. However, the correlation between the driver energy input and the internal energy increase indicates  the presence of wave heating in our domain, similarly to the findings of \citet[][]{mingzhe2023ApJ...949L...1G}.

\section{Discussion and conclusions} \label{sec:discussions}
This aim of this study is to address the effects of random motion footpoint drivers on the  evolution of oscillating loops, with a focus on the manifestation of instabilities. We worked with a straight flux tube, modelling a gravitationally stratified coronal loop with a chromospheric part and an artificially broadened transition region, based on the work done by \citet{pelouze2023A&A...672A.105P} and \citet{mingzhe2023ApJ...949L...1G}. In our analysis, we focused  on the dynamics of the coronal part, while treating the lower chromosphere as a mass reservoir and an anchor to our loop footpoints. We dropped the older approach of a monoperiodic resonant driver, to instead use a coloured noise footpoint driver, similar to the one used in \citet{afanasyev2020decayless}. 

Our drivers excite transverse oscillations in our loop, as a superposition of waves of different frequencies. These oscillations that are observed alongside the background aperiodic signals were shown in the past to have properties of decayless oscillations; specifically, they are standing transverse waves of non-decaying amplitude, with frequencies corresponding to the fundamental standing kink mode and its overtones \citep{Karampelas2024A&A...681L...6K}. Our drivers also excite slow waves, propagating along the loop, which have also been reported in past studies of coronal loops undergoing driven transverse oscillations \citep[e.g.][]{karampelas2019}. 

We report the development of the KH instability in coronal loops  with chromospheric footpoints driven by broadband random motion footpoint drivers. The manifestation and growth of the instability only properly takes place after the onset of the standing waves, after $\sim 1.5$ oscillation periods. Both drivers exhibit an almost identical qualitative behaviour for the spatial and temporal evolution of the KH instability. The only difference is that the instability is slightly more prominent for the red noise driver, which has retained its high power at the lower frequencies, leading to stronger shear velocities.

The KH instability has been reported in the past for transversely oscillating loops, driven by single frequency resonant and non-resonant kink mode drivers \citep[e.g.][]{afanasev2019}, and by broadband transverse wave drivers \citep{Pagano2019,Howson2023Physi...5..140H}. However, only monochromatic transverse wave drivers with frequency equal to that of the fundamental standing kink mode of the oscillating loop had been known to lead to fully deformed cross-sections, as shown in \citet{karampelas2018fd} \citep[see also][for the case of the linearly polarised, monoperiodic resonant driver]{Howson2023Physi...5..140H}. For our set-up of a gravitationally stratified coronal loop with chromospheric footpoints, we showed that broadband transverse wave drivers with a power-law spectrum can also lead to fully turbulent cross-sections due to the KH instability, across the entire volume of the coronal part of the loop. 

In our simulations, the detrended driver has a RMS velocity ($V_{RMS}$) of $\sim 0.94$\,km s$^{-1}$. This is lower than, but comparable to, the RMS velocities of $\sim 1.32$\,km s$^{-1}$ derived from SST and \textit{Hinode} observations of the horizontal motions of solar magnetic bright points \citep{Chitta2012ApJ...752...48C}, which might be regarded as the motion of the loop footpoints. The RMS velocity of the driver is also notably lower than that of harmonic resonant drivers that were used in past studies \citep[for example, $\sim 2.828$\,km s$^{-1}$ in][]{karampelas2019amp}, which are also expected to provide energy more efficiently into the system \citep{Howson2023Physi...5..140H}. Despite these setbacks, our low $V_{RMS}$ driver can still lead to spatially extended KH instability eddies and turbulent coronal parts in our loops. A similar evolution of the KH instability is also reported for our power-law broadband driver with the full red noise spectrum, further supporting the hypothesis that the KH instability should be present in oscillating loops in the solar corona.

Both drivers showcase a small delay in the manifestation of the KH instability near the anchored footpoint with respect to the loop loop apex, of the order of 1 min. This delay can be seen by the initial manifestation of the density maxima in Figure \ref{fig:timedist}. In \citet{depontieu2022ApJ...926...52D...MUSE}, apparent propagation effects, associated with the onset of KH instability, have been reported in synthetic observations with temporal, spatial, and spectral resolution targeted at the upcoming Multi-slit Solar Explorer (MUSE) mission. These propagation effects had been observed only for non-driven impulsively oscillating loops for the fundamental standing kink mode, while they could not be identified for monochromatically and resonantly driven oscillations. Here we have the first report of such apparent propagation effects in simulation data for driven kink oscillations in general, and from broadband drivers with power-law spectra in particular. Our results cast doubt on the potential role of these propagation effects as a tool to distinguish footpoint driven from impulsive transverse loop oscillations, as was suggested in \citet{depontieu2022ApJ...926...52D...MUSE}, bringing the need for further investigation.

As a final point, we need to discuss the potential role of transverse loop oscillations in counterbalancing the optically thin radiative losses, which has been the focus of many recent studies \citep[e.g.][]{mijie2021ApJL,DeMoortel2022ApJ...941...85D}. A crucial piece of the puzzle, is identifying a way for the provided energy to reach the dissipation scales. In \citet{karampelas2019}, the turbulent cascade of energy is   reported in simulations of transversely oscillating coronal loops, fully deformed by the KH instability. In \citet{hillier2020}, the efficiency of KH instability induced turbulence in plasma heating is   debated for decayless loop oscillations driven by a monochromatic resonant driver. However, our current study shows that power-law-type transverse drivers also lead to KH instability-induced turbulence. These drivers also contain a DC, or low-frequency component with a higher power spectral density. The additional energy input from that DC component of the red noise driver leads to a higher temperature increase than the driver without this low-frequency component, similarly to what is discussed in \citet{Howson2022A&A...661A.144H}. The RMS input Poynting flux from the red noise driver in our simulations is calculated to be $\sim 310$\,J\,s$^{-1}$\,m$^{-2}$, as opposed to $\sim 83$\,J\,s$^{-1}$\,m$^{-2}$ for the detrended driver and the bottom boundary. This value for the red noise driver is comparable to the expected radiative losses for the quiet-Sun \citep{withbroenoyes1977ara}. Looking at the base of the corona ($z\sim 7$\,Mm), the corresponding values of the Poynting flux are $\sim 38$\,J\,s$^{-1}$\,m$^{-2}$ for the red noise driver and $\sim 20$\,J\,s$^{-1}$\,m$^{-2}$ for the detrended driver, which are less than the required values to replenish the coronal radiative losses. This is due to the poor transmission of energy across the transition region, which leads to plasma heating and evaporation at the transition region, as shown in Figure \ref{fig:temperature}. We  note here that quantitatively, any numerical results will be affected by a number of different factors, such as the grid resolution in the transition region and the corrections used to resolve the dynamics in the top of the chromosphere. \citet{JohnstonBradshaw2019} showed that using a constant cutoff temperature in the thermal conduction coefficient can lead to an underestimation of coronal densities during evaporation from heating events, while \citet{HowsonBrue2023MNRAS.526..499H} showed that a constant cutoff temperature can lead to an overestimation of the wave energies in the corona. This study also showed that the errors  depend upon the frequencies of the waves that are reaching the transition region, as well as the resolution used. Despite these limitations due to the numerical methods employed, however, the resulting Poynting fluxes at $z=0$ and $z=7$\,Mm can still give us important insight into the evolution of such systems. Due to the small value of the magnetic field near the footpoints in our set-ups, our velocity driver provides a much lower RMS Poynting flux than it would have in a more realistic set-up, with a footpoint magnetic field  of the order of many hundreds G for the same velocity amplitudes. Even if the bulk of that Poynting energy density input does not reach the corona, as shown in Figure \ref{fig:energyinput}, due to attenuation at the transition region \citep[see the cutoff for frequencies below $2$\,mHz in][]{pelouze2023A&A...672A.105P}, it could still lead to heating of the lower atmosphere, and have an indirect effect on the mass and energy balance of oscillating loops, as hinted by our findings for the temperature, mass flux and enthalpy flux in Figure \ref{fig:temperature}.

\begin{acknowledgements}
K.K. acknowledges support by an FWO (Fonds voor Wetenschappelijk Onderzoek – Vlaanderen) postdoctoral fellowship (1273221N). TVD was supported by the European Research Council (ERC) under the European Union's Horizon 2020 research and innovation programme (grant agreement No 724326), the C1 grant TRACEspace of Internal Funds KU Leuven, and a Senior Research Project (G088021N) of the FWO Vlaanderen. M.G. acknowledges support from the National Natural Science Foundation of China (12203030). The computational resources and services used in this work were provided by the VSC (Flemish Supercomputer Center), funded by the Research Foundation Flanders (FWO) and the Flemish Government – department EWI.
\end{acknowledgements}

\bibliographystyle{aa}
\bibliography{paper}

\end{document}